%% file: main.tex
\documentclass[%
floatfix,
twocolumn,%
longbibliography,
author-year,%
author-numerical,%
]{revtex4-1}

\usepackage{amsmath,amssymb}

\usepackage[utf8]{inputenc}
\usepackage{graphicx}

\usepackage[normalem]{ulem} 

\usepackage{type1ec} 

\usepackage[outer=1.3cm, inner = 1.3cm, top= 1.8cm, bottom = 1.8cm]{geometry}

\usepackage{color}
\usepackage[dvipsnames]{xcolor}

\input{commands}

\usepackage{setspace}

\begin{document}

\title{Towards an analytical description of active microswimmers in clean and in surfactant-covered drops}


\thanks{Regular article  contributed  to  the  Topical  Issue  of  the  European Physical Journal~E  entitled ``Physics of Motile Active Matter'' edited by Gerhard Gompper, Clemens Bechinger, Holger Stark, and Roland G.\ Winkler, in the frame of the Priority Program ``Microswimmers -- from single particle motion to collective behavior'' (SPP1726) funded by the Deutsche Forschungsgemeinschaft (DFG).}



\author{Alexander R.\ Sprenger$^1$}
\email{sprenger@thphy.uni-duesseldorf.de}
\affiliation
{$^1$Institut f\"{u}r Theoretische Physik II: Weiche Materie, Heinrich-Heine-Universit\"{a}t D\"{u}sseldorf, Universit\"{a}tsstra\ss e 1, D-40225 D\"{u}sseldorf, Germany \\
$^2$School of Mechanical Engineering, Purdue University, West Lafayette, IN 47907, USA \\
$^3$Institute of Theoretical Physics, Faculty of Physics, University of Warsaw, Pasteura 5, 02-093 Warsaw, Poland \\
$^4$Department of Bioengineering, Stanford University, 443 Via Ortega, Stanford, CA 94305, USA \\
$^5$Centro de Investigaci{\'o}n DAiTA Lab, Facultad de  Estudios Interdisciplinarios, Universidad Mayor, Av.\  Manuel  Montt  367,  Providencia,  Santiago de Chile, Chile \\
$^6$Institut f\"{u}r Physik, Otto-von-Guericke-Universit\"{a}t Magdeburg, Universit\"{a}tsplatz 2, 39106 Magdeburg, Germany}

\author{Vaseem A.\ Shaik$^2$}
\affiliation
{$^1$Institut f\"{u}r Theoretische Physik II: Weiche Materie, Heinrich-Heine-Universit\"{a}t D\"{u}sseldorf, Universit\"{a}tsstra\ss e 1, D-40225 D\"{u}sseldorf, Germany \\
$^2$School of Mechanical Engineering, Purdue University, West Lafayette, IN 47907, USA \\
$^3$Institute of Theoretical Physics, Faculty of Physics, University of Warsaw, Pasteura 5, 02-093 Warsaw, Poland \\
$^4$Department of Bioengineering, Stanford University, 443 Via Ortega, Stanford, CA 94305, USA \\
$^5$Centro de Investigaci{\'o}n DAiTA Lab, Facultad de  Estudios Interdisciplinarios, Universidad Mayor, Av.\  Manuel  Montt  367,  Providencia,  Santiago de Chile, Chile \\
$^6$Institut f\"{u}r Physik, Otto-von-Guericke-Universit\"{a}t Magdeburg, Universit\"{a}tsplatz 2, 39106 Magdeburg, Germany}

\author{Arezoo M.\ Ardekani$^2$}
\affiliation
{$^1$Institut f\"{u}r Theoretische Physik II: Weiche Materie, Heinrich-Heine-Universit\"{a}t D\"{u}sseldorf, Universit\"{a}tsstra\ss e 1, D-40225 D\"{u}sseldorf, Germany \\
$^2$School of Mechanical Engineering, Purdue University, West Lafayette, IN 47907, USA \\
$^3$Institute of Theoretical Physics, Faculty of Physics, University of Warsaw, Pasteura 5, 02-093 Warsaw, Poland \\
$^4$Department of Bioengineering, Stanford University, 443 Via Ortega, Stanford, CA 94305, USA \\
$^5$Centro de Investigaci{\'o}n DAiTA Lab, Facultad de  Estudios Interdisciplinarios, Universidad Mayor, Av.\  Manuel  Montt  367,  Providencia,  Santiago de Chile, Chile \\
$^6$Institut f\"{u}r Physik, Otto-von-Guericke-Universit\"{a}t Magdeburg, Universit\"{a}tsplatz 2, 39106 Magdeburg, Germany}

\author{Maciej Lisicki$^3$}
\affiliation
{$^1$Institut f\"{u}r Theoretische Physik II: Weiche Materie, Heinrich-Heine-Universit\"{a}t D\"{u}sseldorf, Universit\"{a}tsstra\ss e 1, D-40225 D\"{u}sseldorf, Germany \\
$^2$School of Mechanical Engineering, Purdue University, West Lafayette, IN 47907, USA \\
$^3$Institute of Theoretical Physics, Faculty of Physics, University of Warsaw, Pasteura 5, 02-093 Warsaw, Poland \\
$^4$Department of Bioengineering, Stanford University, 443 Via Ortega, Stanford, CA 94305, USA \\
$^5$Centro de Investigaci{\'o}n DAiTA Lab, Facultad de  Estudios Interdisciplinarios, Universidad Mayor, Av.\  Manuel  Montt  367,  Providencia,  Santiago de Chile, Chile \\
$^6$Institut f\"{u}r Physik, Otto-von-Guericke-Universit\"{a}t Magdeburg, Universit\"{a}tsplatz 2, 39106 Magdeburg, Germany}

\author{Arnold J. T. M. Mathijssen$^4$}
\affiliation
{$^1$Institut f\"{u}r Theoretische Physik II: Weiche Materie, Heinrich-Heine-Universit\"{a}t D\"{u}sseldorf, Universit\"{a}tsstra\ss e 1, D-40225 D\"{u}sseldorf, Germany \\
$^2$School of Mechanical Engineering, Purdue University, West Lafayette, IN 47907, USA \\
$^3$Institute of Theoretical Physics, Faculty of Physics, University of Warsaw, Pasteura 5, 02-093 Warsaw, Poland \\
$^4$Department of Bioengineering, Stanford University, 443 Via Ortega, Stanford, CA 94305, USA \\
$^5$Centro de Investigaci{\'o}n DAiTA Lab, Facultad de  Estudios Interdisciplinarios, Universidad Mayor, Av.\  Manuel  Montt  367,  Providencia,  Santiago de Chile, Chile \\
$^6$Institut f\"{u}r Physik, Otto-von-Guericke-Universit\"{a}t Magdeburg, Universit\"{a}tsplatz 2, 39106 Magdeburg, Germany}

\author{Francisca Guzm\'{a}n-Lastra$^5$}
\affiliation
{$^1$Institut f\"{u}r Theoretische Physik II: Weiche Materie, Heinrich-Heine-Universit\"{a}t D\"{u}sseldorf, Universit\"{a}tsstra\ss e 1, D-40225 D\"{u}sseldorf, Germany \\
$^2$School of Mechanical Engineering, Purdue University, West Lafayette, IN 47907, USA \\
$^3$Institute of Theoretical Physics, Faculty of Physics, University of Warsaw, Pasteura 5, 02-093 Warsaw, Poland \\
$^4$Department of Bioengineering, Stanford University, 443 Via Ortega, Stanford, CA 94305, USA \\
$^5$Centro de Investigaci{\'o}n DAiTA Lab, Facultad de  Estudios Interdisciplinarios, Universidad Mayor, Av.\  Manuel  Montt  367,  Providencia,  Santiago de Chile, Chile \\
$^6$Institut f\"{u}r Physik, Otto-von-Guericke-Universit\"{a}t Magdeburg, Universit\"{a}tsplatz 2, 39106 Magdeburg, Germany}

\author{Hartmut L\"{o}wen$^1$}
\affiliation
{$^1$Institut f\"{u}r Theoretische Physik II: Weiche Materie, Heinrich-Heine-Universit\"{a}t D\"{u}sseldorf, Universit\"{a}tsstra\ss e 1, D-40225 D\"{u}sseldorf, Germany \\
$^2$School of Mechanical Engineering, Purdue University, West Lafayette, IN 47907, USA \\
$^3$Institute of Theoretical Physics, Faculty of Physics, University of Warsaw, Pasteura 5, 02-093 Warsaw, Poland \\
$^4$Department of Bioengineering, Stanford University, 443 Via Ortega, Stanford, CA 94305, USA \\
$^5$Centro de Investigaci{\'o}n DAiTA Lab, Facultad de  Estudios Interdisciplinarios, Universidad Mayor, Av.\  Manuel  Montt  367,  Providencia,  Santiago de Chile, Chile \\
$^6$Institut f\"{u}r Physik, Otto-von-Guericke-Universit\"{a}t Magdeburg, Universit\"{a}tsplatz 2, 39106 Magdeburg, Germany}

\author{Andreas M.\ Menzel$^6$}
\affiliation
{$^1$Institut f\"{u}r Theoretische Physik II: Weiche Materie, Heinrich-Heine-Universit\"{a}t D\"{u}sseldorf, Universit\"{a}tsstra\ss e 1, D-40225 D\"{u}sseldorf, Germany \\
$^2$School of Mechanical Engineering, Purdue University, West Lafayette, IN 47907, USA \\
$^3$Institute of Theoretical Physics, Faculty of Physics, University of Warsaw, Pasteura 5, 02-093 Warsaw, Poland \\
$^4$Department of Bioengineering, Stanford University, 443 Via Ortega, Stanford, CA 94305, USA \\
$^5$Centro de Investigaci{\'o}n DAiTA Lab, Facultad de  Estudios Interdisciplinarios, Universidad Mayor, Av.\  Manuel  Montt  367,  Providencia,  Santiago de Chile, Chile \\
$^6$Institut f\"{u}r Physik, Otto-von-Guericke-Universit\"{a}t Magdeburg, Universit\"{a}tsplatz 2, 39106 Magdeburg, Germany}

\author{Abdallah Daddi-Moussa-Ider$^1$}
\email{ider@thphy.uni-duesseldorf.de}
\affiliation
{$^1$Institut f\"{u}r Theoretische Physik II: Weiche Materie, Heinrich-Heine-Universit\"{a}t D\"{u}sseldorf, Universit\"{a}tsstra\ss e 1, D-40225 D\"{u}sseldorf, Germany \\
$^2$School of Mechanical Engineering, Purdue University, West Lafayette, IN 47907, USA \\
$^3$Institute of Theoretical Physics, Faculty of Physics, University of Warsaw, Pasteura 5, 02-093 Warsaw, Poland \\
$^4$Department of Bioengineering, Stanford University, 443 Via Ortega, Stanford, CA 94305, USA \\
$^5$Centro de Investigaci{\'o}n DAiTA Lab, Facultad de  Estudios Interdisciplinarios, Universidad Mayor, Av.\  Manuel  Montt  367,  Providencia,  Santiago de Chile, Chile \\
$^6$Institut f\"{u}r Physik, Otto-von-Guericke-Universit\"{a}t Magdeburg, Universit\"{a}tsplatz 2, 39106 Magdeburg, Germany}

\date{\today}

\begin{abstract}
	Geometric confinements are frequently encountered in the biological world and strongly affect the stability, topology, and transport properties of active suspensions in viscous flow.
	Based on a far-field analytical model, the low-Reynolds-number locomotion of a self-propelled microswimmer moving inside a clean viscous drop or a drop covered with a homogeneously distributed surfactant, is theoretically examined.
	The interfacial viscous stresses induced by the surfactant are described by the well-established Boussinesq-Scriven constitutive rheological model.
	Moreover, the active agent is represented by a force dipole and the resulting fluid-mediated hydrodynamic couplings between the swimmer and the confining drop are investigated.
	We find that the presence of the surfactant significantly alters the dynamics of the encapsulated swimmer by enhancing its reorientation.
	Exact solutions for the velocity images for the Stokeslet and dipolar flow singularities inside the drop are introduced and expressed in terms of infinite series of harmonic components.
	Our results offer useful insights into guiding principles for the control of confined active matter systems and support the objective of utilizing synthetic microswimmers to drive drops for targeted drug delivery applications.	
\end{abstract}
\maketitle

\section{Introduction}

Controlled locomotion of nano- and microscale objects in viscous media is of considerable importance in many areas of engineering and science~\cite{gompper20}. 
Synthetic nano- and micro-motors hold significant promise for future biotechnological and medical applications such as precise assembly of materials~\cite{palacci13, sacanna13, chronis05, junkin11, diller14, ahmed15, schmidt19}, non-invasive microsurgery~\cite{kagan12, walker15, nelson10}, targeted drug delivery~\cite{qiu15, park17, mostaghaci17, gourevich13, kei14}, and biosensing~\cite{wang88}.
Over the last few decades, there has been a rapidly mounting interest among researchers in understanding and unveiling the physics of self-propelled active particles and microswimmers, see Refs.~\onlinecite{lauga09,marchetti13, elgeti15, menzel15, bechinger16, zottl16, lauga16ARFM, illien17, desai2017modeling, dey19, shaebani20} for recent reviews.
Various intriguing effects of collective behavior are displayed and fascinating self-organized spatiotemporal patterns are created by the mutual interaction of many active agents.
Notable examples include the formation of propagating density waves~\cite{gregoire04, mishra10, menzel12}, the emergence of mesoscale turbulence~\cite{wensink12, wensink12pnas, dunkel13, grossmann14, kaiser14, heidenreich14, heidenreich16, doostmohammadi17}, the motility-induced phase separation~\cite{cates15, cates13, tailleur08, buttinoni13, speck14, speck15, digregorio18, mandal19, caprini20}, and lane formation~\cite{menzel13, kogler15, romanczuk16, menzel16njp, reichhardt18, wachtler2016lane, klapp2016collective}.

In many biologically and technologically relevant situations, actively swimming biological microorganisms and artificial self-driven particles are present. Typically, they function and survive in confined environments which are known to strongly affect their swimming and propulsion behavior as well as the transport properties in viscous media.
Examples include \textit{Bacillus subtilis} in soil~\cite{earl08, pantastico92}, \textit{Escherichia coli} in intestines~\cite{moon79, palmela18}, pathogenic bacteria in microvasculature~\cite{Moriarty:2008}, and spermatozoa navigation through the mammalian female reproductive tract~\cite{suarez06, kantsler14, Nosrati:2015}.
Geometric confinement caused by a plane rigid or fluid interface affects the dynamics of microswimmers by altering their speed and orientation with respect to the interface~\cite{Reynolds1965, Katz1974,Berke2008, elgeti09, Li2009,llopis10, Spagnolie2012, ishimoto13, ishimoto14, li14, Lopez2014, uspal15, ibrahim15, schaar15, das15, elgeti16, mozaffari16, simmchen16, lintuvuori16, lushi17, ishimoto17, yazdi17, ruhle18, mozaffari18, shen18, mirzakhanloo18, ahmadzadegan2019hydrodynamic} and changing their swimming trajectories from straight lines in a bulk fluid to circular shapes near interfaces~\cite{Lauga2006, DiLeonardo2011,hu15, daddi2018swimming, daddi2018hydrodynamic, mousavi20}.
Studies of the dynamics of microswimmers in a microchannel bounded by two interfaces~\cite{bilbao13, wu15, wu16, daddi2018state, dalal20} or immersed in a thin liquid film~\cite{brotto13, Mathijssen2016, Desai2020} or spherical cavity~\cite{das2019dynamics} revealed complex evolution scenarios of microswimmers in the presence of narrow confinement~\cite{malgaretti19}.

Curved boundaries strongly affect the stability and topology of active suspensions under confinement and drive self-organization in a wide class of active matter systems~\cite{wioland13, theillard17, gao17}.
For instance, a dense aqueous suspension of \textit{Bacillus subtilis} confined inside a viscous drop self-organizes into a stable spiral vortex surrounded by a counter-rotating boundary layer of motile cells~\cite{wioland13, Lushi2014}.
In addition, a sessile drop containing photocatalytic particles exhibits a transition to a collective behavior leading to self-organized flow patterns~\cite{singh20}.
Under the effect of an external magnetic field, swimming magnetotactic bacteria confined into water-in-oil drops can self-assemble into a rotary motor that exerts a net torque on the surrounding oil phase~\cite{Vincenti2019, Ramos2020}.
In microfluidic systems, synthetic microswimmers, such as artificial bacterial flagella, are frequently used to drive drops in the context of targeted drug delivery systems~\cite{mirkovic10, Ding2016}. 
Along these lines, nontrivial dynamics of a particle-encapsulating drop in shear flow were revealed~\cite{zhu17}.
To understand the self-organization or the energy transport from the swimmer scale to the system scale or to develop efficient and reliable drug delivery systems, we need to unravel the physics underlying the dynamics of a motile microorganism encapsulated inside drop.
This is the focus of the present work, concentrating on clean drops or those covered by a surfactant.

The swimming dynamics in the vicinity of a rigid spherical obstacle~\cite{Takagi2014, Spagnolie2015, Jashnsaz2017}, a clean or a surfactant-covered drop~\cite{Desai2018, Desai2018a, Desai2019} have been investigated theoretically.
It has been demonstrated that a swimming organism reorients itself and gets scattered from the obstacle or gets trapped or captured by it if the size of the obstacle is large enough and the settling/rising speed of the microorganism is small enough. 
Near a viscous drop, the surfactant increases the trapping capability~\cite{Desai2018} and can even break the kinematic reversibility associated with the inertialess realm of swimming microorganisms~\cite{Shaik2018}.
In contrast to that, the presence of a surfactant near a planar interface was found not to change the reorientation dynamics~\cite{Lopez2014} but to change the swimming speed~\cite{Shaik2019} in addition to the circling direction~\cite{Lopez2014}.

In the theoretical investigation of locomotion under confinement, swimming microorganisms are commonly approximated by microswimmer models, frequently using a far-field representation based on higher-order flow singularities~\cite{lauga09}. 
Well-established model microswimmers include Taylor's swimming sheet~\cite{Taylor1951, elfring09, sauzade11, dasgupta13, montenegro14} and the spherical squirmer~\cite{Lighthill1952, Blake1971,pak14, gotze10, thutupalli11, zhu12, zhu13, zottl14, lintuvuori17, kuhr17, zottl18, kuron2019lattice, kuhr19, zottl20}.
The former is a good representation of the tail of human spermatozoa and \textit{Caenorhabditis elegans} while the latter is believed to describe well the behavior of \textit{Paramecium}, \textit{Opalina}, and \textit{Volvox}. 
Linked spheres that are able to propel forward when the mutual distance between the spheres is varied in a nonreciprocal fashion constitute another class of model
microswimmers~\cite{najafi04, golestanian08, golestanian08epje,liebchen18, alexander09, ziegler19, sukhov19, nasouri19}.
Moreover, various minimal model microswimmers have been proposed to model swimming agents with rigid bodies and flexible propelling appendages~\cite{menzel16,hoell17, adhyapak17, adhyapak2018ewald, hoell18, daddi18nematic, daddi2020dynamics, hoell19}.
Many of the organisms are approximately neutrally buoyant, so they hardly experience any gravitational force or torque. 
This implies that the action of a swimming organism in far-field representation can conveniently be described by a force dipole and higher-order singularities to investigate its motion under confinement. 
The accuracy of this simple far-field analysis was verified by comparison with other theoretical and fully resolved computer simulations~\cite{Spagnolie2012, Mathijssen2016, Shaik2017}. 
In particular, the far-field analysis was shown to predict and reproduce experimental and numerical observations~\cite{Berke2008, Lopez2014, Desai2020, Spagnolie2015}. 
This motivates us to employ the far-field representation to examine the swimming behavior inside a clean or surfactant-covered viscous drop.

Theoretically, one of the first studies of low-Reynolds-number locomotion inside a drop considered a spherical squirmer encapsulated inside a drop of a comparable size immersed in an otherwise quiescent viscous medium~\cite{Reigh2017, reigh17prf}. The analytical theory was complemented and supplemented by numerical implementations based on a boundary element method~\cite{pozrikidis92book}.
It was reported that the drop can be propelled by the encaged swimmer, and in some situations the swimmer-drop composite remains in a stable co-swimming state so that the swimmer and drop maintain a concentric configuration and move with the same velocity~\cite{Reigh2017}.
Meanwhile, the presence of a surfactant on the surface of the drop was found to increase or decrease the squirmer or drop velocities depending on the precise location of the swimmer inside the confining drop~\cite{Shaik2018}.
In the presence of a shear flow, it was demonstrated that the activity of a squirmer inside a drop can significantly enhance or reduce the deformation of the drop depending on the orientation of the swimmer~\cite{K.V.S.2020}.  
More recently, the dynamics of a drop driven by an internal active device composed of a three-point-force moving on a prescribed track was examined~\cite{Rueckert2019, kree20}.

The dynamics of a squirmer inside a drop is not analytically tractable for arbitrary positions and orientations of the swimmer. Therefore, recourse to numerical techniques is generally necessary to obtain a complete understanding of the low-Reynolds-number locomotion~\cite{Reigh2017}. 
However, when keeping all details, these methods are not easily extensible to the case of multiple swimmers. 
To deal with these limitations, the swimming organism can be modeled in the far-field limit under confinement using the classical method of images~\cite{blake71, swan07}. 
The latter has the advantage of being easily extensible to the case of a drop containing many active and hydrodynamically interacting organisms in the dilute suspension limit.
In this context, an image system for a point force bounded by a rigid spherical container has previously been reported~\cite{usha93, Maul1994, Maul1996, hasimoto97, sellier08, felderhof12cavity, aponte16, aponte17}.
Nevertheless, image systems for force dipoles or higher-order singularities bounded by a spherical fluid interface possibly covered by a surfactant are still missing.

In the present contribution, we derive the image solution for a point force (Stokeslet) and dipole singularities inside a spherical drop, both with a clean surface, or covered by a surfactant. 
We model the interfacial viscous stresses at the surfactant-covered drop boundary by the well-established Boussinesq-Scriven rheological constitutive model~\cite{brenner13}.
Our approach is based on the method originally introduced by Fuentes \textit{et al.}~\cite{fuentes88, fuentes89}, who derived the solution for a Stokeslet acting outside a clean viscous drop.
An analogous approach was employed by some of us to derive the Stokeslet solution near~\cite{daddi17b, daddi17c} or inside~\cite{daddi18creeping, hoell19creeping} a spherical elastic object, and outside a surfactant-covered drop~\cite{shaik17}.
We find that the presence of the surfactant alters the swimming behavior of the encaged microswimmer by enhancing its rate of rotation.

We organize the remainder of the paper as follows.
In Sec.~\ref{sec:Mono}, we derive the solution for the viscous flow field induced by an axisymmetric or symmetric Stokeslet acting inside a clean and a surfactant-covered drop.
We then use this flow field in Sec.~\ref{sec:Dipo} to obtain the corresponding image solution for a force-dipole singularity of arbitrary location and orientation within the spherical drop.
In Sec.~\ref{sec:SwimmerDynamics}, we derive the induced translational and rotational velocities resulting from hydrodynamic couplings in the present geometry.
Finally, concluding remarks are contained in Sec.~\ref{sec:Conclusions} and technical details are shifted to Appendices~\ref{appendix:1} and~\ref{appendix:2}.

\section{Monopole singularity}\label{sec:Mono}

\begin{figure}
	\centering
	\includegraphics[scale=1.2]{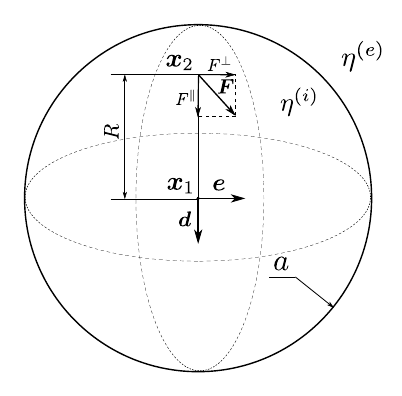}
	\caption{Schematic illustration of the system setup. A point-force singularity of strength~$\vect{F}$ is acting at the position~$\vect{x}_2$ inside a spherical viscous drop of radius~$a$.
		The origin of the system of coordinates coincides with the center of the drop~$\vect{x}_1$.
		We denote the distance between the origin and the position of the singularity as~$R$.
		The viscosities inside and outside the drop are designated as~$\eta^{(i)}$ and~$\eta^{(e)}$, respectively.
		For an arbitrary orientation, the point force is decomposed into an axisymmetric component~$F^\parallel$ directed along the unit vector~$\vect{d}$ and an transverse component~$F^\perp$ pointing along the unit vector~$\vect{e}$.
		Without loss of generality, the point force is taken to be located on the $z$ axis, with components along $z$ and $x$ directions, where $\vect{d}=-\vect{\hat{z}}$ and	$\vect{e}=\vect{\hat{x}}$.
	}\label{Fig:system}
\end{figure}

We derive the solution of the viscous incompressible flow induced by a point-force singularity of strength~$\vect{F}$ acting at position~$\vect{x}_2$ inside a viscous drop of radius~$a$.
The origin of the system of coordinates is located at position~$\vect{x}_1$, the center of the viscous drop.
We denote by~$\R = \vect{x} - \vect{x}_1$ the position vector and by $r := \left| \R \right|$ the radial distance from the origin.
Moreover, we refer by $\eta^{(i)}$ and~$\eta^{(e)}$ to the dynamic viscosities of the Newtonian fluids inside and outside the drop, respectively.
Next, we define the unit vector $\vect{d} = \left( \vect{x}_1 - \vect{x}_2 \right)/R$ with $R = \left| \vect{x}_1 - \vect{x}_2 \right|$ denoting the distance between the singularity position and the origin. 
In addition, we define the unit vector~$\vect{e}$ orthogonal to~$\vect{d}$ so that the force~$\vect{F}$ can be decomposed into an axisymmetric component $F^\parallel \vect{d}$ and a transverse component $F^\perp \vect{e}$.
See Fig.~\ref{Fig:system} for an illustration of the system setup.

In the remainder of this article, we rescale all lengths by the radius~$a$ of the drop. We will denote by superscripts~$(i)$ and~$(e)$ quantities referring to the inside and outside of the drop, respectively. The problem of finding the incompressible hydrodynamic flow is thus equivalent to solving the singularly forced Stokes equations~\cite{kim13} for the fluid inside the drop,
\begin{subequations}\label{StokesInside}
	\begin{align}
		\eta^{(i)} \boldsymbol{\nabla}^2 \vect{v}^{(i)} - \boldsymbol{\nabla} p^{(i)} + \vect{F} \delta \left( \vect{x} - \vect{x}_2 \right) &= \vect{0} \, , \\
			\boldsymbol{\nabla} \cdot \vect{v}^{(i)}         &= 0
	\end{align}
\end{subequations}
for $r<1$, and the homogeneous Stokes equations for the fluid outside the drop,
\begin{subequations}\label{StokesOutside}
	\begin{align}
		\eta^{(e)} \boldsymbol{\nabla}^2 \vect{v}^{(e)} - \boldsymbol{\nabla} p^{(e)} &= \vect{0} \, , \\
		\boldsymbol{\nabla} \cdot \vect{v}^{(e)} &= 0
	\end{align}	
\end{subequations}
for $r>1$, wherein~$\vect{v}^{(q)}$ and~$p^{(q)}, q \in \{i,e\}$, denote the corresponding fluid-velocity and pressure fields, respectively. 
We focus on the small-deformation regime concerning the shape of the drop so that deviations from sphericity are assumed to be negligible.
Moreover, we first assume the drop to be stationary. This implies that it is held fixed in space, for instance by means of optical tweezers~\cite{neuman04}.
Accordingly, the radial component of the fluid-velocity field at the surface of the stationary drop is assumed to vanish in the frame of reference associated with the viscous drop.

Under these assumptions, Eqs.~\eqref{StokesInside} and~\eqref{StokesOutside} are subject to the regularity conditions 
\begin{equation}
		|\vect{v}^{(i)}| < \infty \text{~~~for~~~} r\to 0 \, , \quad
		\vect{v}^{(e)} \to \mathbf{0} \text{~~~as~~~} r\to \infty \, , \label{regularityCond}
\end{equation}
in addition to the boundary conditions imposed at the surface of the stationary drop at~$r=1$,
\begin{subequations}\label{boundaryCond1}
	\begin{align}
		v_r^{(i)} = v_r^{(e)} &= 0   \, , \label{VanishingRadial} \\
		\vect{v}_\mathrm{S} := \boldsymbol{\Pi} \cdot \vect{v}^{(i)}
		&= \boldsymbol{\Pi} \cdot \vect{v}^{(e)}   \, , \label{ContVelo} 
	\end{align}
\end{subequations}
where $\boldsymbol{\Pi} = \boldsymbol{1} - \vect{e}_r \vect{e}_r$ is the projection operator, with~$\boldsymbol{1}$ denoting the identity tensor, and $\vect{v}_\mathrm{S}$ is the tangential velocity.
Equation~\eqref{VanishingRadial} represents the kinematic condition stating that the drop remains undeformed whereas Eq.~\eqref{ContVelo} stands for the natural continuity of the tangential velocities across the surface of the drop.

On the one hand, for a clean drop, \textit{i.e.}, without surfactant, shear elasticity, or bending rigidity, the tangential hydrodynamic stresses across the surface of the drop are continuous~\cite{felderhof89creeping}. 
Then,
\begin{equation} \label{boundaryCond2clean}
	\boldsymbol{\Pi} \cdot \left( \vect{T}^{(i)} - \vect{T}^{(e)}  \right) = \vect{0} \, , 
\end{equation}
where $\vect{T}^{(q)} = \boldsymbol{\sigma}^{(q)} \cdot \vect{e}_r$ with $\boldsymbol{\sigma}^{(q)} $, $q \in \{i,e\}$, denoting the hydrodynamic viscous stress tensor.

On the other hand, to model the surfactant-covered drop, we use the boundary conditions~\cite{shaik17}
\begin{subequations}\label{boundaryCond2surfactant}
	\begin{align}
		\boldsymbol{\nabla}_\mathrm{S} \cdot \vect{v}_\mathrm{S} &= 0 \, , \label{IncompressSurfactant} \\ 
		\boldsymbol{\Pi} \cdot \left( \vect{T}^{(i)} - \vect{T}^{(e)}  \right)
		&= \boldsymbol{\nabla}_\mathrm{S} \gamma 
		+ \boldsymbol{\nabla}_\mathrm{S} \cdot \boldsymbol{\tau}_\mathrm{S} \, , \label{Boussinesq} 
	\end{align}
\end{subequations}
where $\gamma$ denotes the interfacial tension, $\boldsymbol{\nabla}_\mathrm{S} = \boldsymbol{\Pi} \cdot \boldsymbol{\nabla}$ is the surface gradient operator, and $\boldsymbol{\tau}_\mathrm{S}$ is the interfacial viscous stress tensor.
Using the Boussinesq-Scriven constitutive law we have~\cite{shaik17}
\begin{equation}
	\boldsymbol{\nabla}_\mathrm{S} \cdot \boldsymbol{\tau}_\mathrm{S} 
	= \eta_\mathrm{S} \left( \frac{2 \vect{v}_\mathrm{S}}{r^2} 
	+ \frac{1}{r \sin\theta} \frac{\partial \varpi}{\partial \phi} \, \vect{e}_\theta
	- \frac{1}{r} \frac{\partial \varpi}{\partial \theta} \, \vect{e}_\phi \right) \, , 
\end{equation}
wherein $\theta$ and $\phi$, respectively, denote the polar and azimuthal angles in the system of spherical coordinates attached to the center of the drop, $\eta_\mathrm{S}$ denotes the interfacial shear viscosity, which we assume to be constant, and 
\begin{equation}
	\varpi = \frac{1}{r\sin\theta} 
	\left( \frac{\partial v_\theta}{\partial \phi}
	- \frac{\partial}{\partial \theta} \left( v_\phi \sin\theta \right) \right) \, .
\end{equation}
Equation~\eqref{IncompressSurfactant} represents the transport equation for an insoluble, non-diffusing, incompressible, and homogeneously distributed surfactant~\cite{stone90, shaik17}, which may be rewritten as
\begin{equation}
	\frac{\partial {v_\mathrm{S}}_\phi}{\partial \phi} + \frac{\partial}{\partial \theta}
	\left( {v_\mathrm{S}}_\theta \sin\theta \right) = 0 \, . \label{gradSVs2}
\end{equation}
We note that the tangential components of the viscous stress vector are expressed in the usual way as
\begin{subequations}
	\begin{align}
		T_{\theta}^{(q)} &= \eta^{(q)} \left( \frac{\partial v_\theta^{(q)}}{\partial r} + \frac{1}{r} \left( \frac{\partial v_r^{(q)}}{\partial \theta} - v_\theta^{(q)} \right) \right) \, , \\
		T_{\phi}^{(q)} &= \eta^{(q)} \left( \frac{\partial v_\phi^{(q)}}{\partial r} + \frac{1}{r} \left( \frac{1}{\sin\theta} \frac{\partial v_r^{(q)}}{\partial \phi} - v_\phi^{(q)} \right) \right) \, , 
	\end{align}
\end{subequations}
for $q \in \{i, e\}$.

To solve the Stokes equations \eqref{StokesInside} and \eqref{StokesOutside}, we write the solution for the fluid flow inside the drop as a sum of two contributions
\begin{equation}
 	\vect{v}^{(i)} = \vStok + \vect{v}^* \, , 
\end{equation}
wherein $\vStok$ denotes the velocity field induced by a point-force singularity in an unbounded bulk medium of viscosity~$\eta^{(i)}$, \textit{i.e.}, in the case of an infinitely extended drop, and $\vect{v}^*$ is the auxiliary solution (also known as the image or reflected flow field) that is required to satisfy the above regularity and boundary conditions.

We now sketch briefly the main steps of the resolution procedure.
First, the fluid velocity induced by the free-space Stokeslet~$\vStok$ for an infinitely large drop is expressed in terms of harmonic functions based at~$\xTwo$, which are subsequently transformed into harmonics based at~$\xOne$ by means of the Legendre expansion~\cite{andrews98}.
Second, the image solution $\vect{v}^{*}$ as well as the flow field outside the cavity~$\vect{v}^{(e)}$ are, respectively, expressed in terms of interior and exterior harmonics based at~$\xOne$.
To this end, we make use of Lamb's general solution of Stokes flows in a spherical domain~\cite{lamb32, cox69, happel12}.
Finally, the unknown series expansion coefficients associated with each fluid domain are determined by satisfying the boundary conditions prescribed at the surface of the drop.

Thanks to the linearity of the Stokes equations, the Green's function for a point force directed along an arbitrary direction in space can be obtained by linear superposition of the solutions for the axisymmetric and transverse problems~\cite{Maul1996}. 
In the following, we detail the derivation of the solution for these two problems independently.

\subsection{Axisymmetric Stokeslet}

The velocity field induced by a free-space Stokeslet located at~$\vect{x}_2$ is expressed in terms of the Oseen tensor as
\begin{equation}
	\vect{v}^\mathrm{S} =  \boldsymbol{\mathcal{G}} (\vect{x} - \vect{x}_2 ) \cdot \vect{F} 
	= \frac{1}{8\pi\eta^{(i)}} 
	\left( \frac{\boldsymbol{1}}{s} + \vect{s} \boldNabla_2 \left( \frac{1}{s} \right)  \right) \cdot \vect{F}  \, , 
\end{equation}
where $\vect{s} = \vect{x} - \vect{x}_2$, $s = \left| \vect{s} \right|$, and $\boldNabla_2 = \partial / \partial \vect{x}_2$ stands for the partial derivative with respect to the singularity position.
The details of derivation have previously been reported by some of us in Ref.~\onlinecite{daddi18creeping}, and will thus be omitted here.
As shown there, the free-space Stokeslet for an axisymmetric point force $\vect{F} = F^\parallel \vect{d}$ can be expanded in terms of an infinite series of harmonics centered at~$\vect{x}_1$ via the Legendre expansion as
\begin{equation}
	8\pi\eta^{(i)} \vStok = F^\parallel \infSumOne  \Bigg( \alpha_n \boldNabla \varphi_n 
    -\frac{2(n+1)}{2n-1} \, \vect{r} \varphi_n \Bigg) R^{n-1} \, , \label{StokesletAxi}
\end{equation}
wherein
\begin{equation}
	\alpha_n = \frac{n-2}{2n-1} \, r^2  - \frac{n}{2n+3} \, R^{2} \, , 
\end{equation}
and~$\varphi_n$ are harmonics of degree~$n$, that are related to the Legendre polynomials of degree~$n$ via~\cite{abramowitz72}
\begin{equation}
 \varphi_n (r, \theta) = \frac{(\vecD \cdot \boldNabla)^n}{n!} \frac{1}{r} = r^{-(n+1)} \, P_n (\cos \theta) \, .
\end{equation}

In addition, the image solution inside the drop can readily be determined from Lamb's general solution~\cite{cox69, misbah06}, and can conveniently be expressed in terms of \emph{interior} harmonics based at $\xOne$ as~\cite{daddi18creeping} 
\begin{equation}
  8\pi\eta^{(i)} \vect{v}^{*} = F^\parallel \sum_{n=1}^{\infty} \left( 
           A_n^\parallel \vect{S}_n^{(1)} + B_n^\parallel  \vect{S}_n^{(2)}
		  \right)  \, , \label{ImageSolutionAxi}
\end{equation}
where we have defined the vector functions
\begin{subequations} \label{Sn}
	\begin{align}
		\vect{S}_n^{(1)} &= \frac{1}{2} \bigg( \left(n+3\right) r^{2} \, \boldNabla \varphi_n  
		                  +  (n+1) (2n+3) \R \varphi_n \bigg) r^{2n+1} \, , \\
		\vect{S}_n^{(2)} &= \bigg( r^{2} \boldNabla \varphi_n + (2n+1) \R \varphi_n \bigg) r^{2n-1} \, .
	\end{align}
\end{subequations}

The total flow field inside the drop is obtained by summing both contributions stated by Eqs.~\eqref{StokesletAxi} and~\eqref{ImageSolutionAxi}, while the series coefficients $A_n^\parallel$ and $B_n^\parallel$ remain to be determined.

Next, the solution of the flow problem outside the drop can likewise be obtained using Lamb's general solution, and can be expressed in terms of \emph{exterior} harmonics based at $\xOne$ as~\cite{daddi18creeping}
\begin{equation}
  8\pi\eta^{(i)} \vect{v}^{(e)} = F^\parallel \sum_{n=1}^{\infty} \Bigg( 
          a_n^\parallel \boldsymbol{\Phi}  
          + b_n^\parallel \boldNabla \varphi_n \Bigg) \, ,
             \label{OutsideSolutionAxi}
\end{equation}
where
\begin{equation}
	\boldsymbol{\Phi}   = (n+1) \, \R \varphi_n -\frac{n-2}{2} \, r^2 \, \boldNabla \varphi_n \, . 
\end{equation}
It is worth noting that, for the ease of matching the boundary conditions at the surface of the drop, we have chosen to rescale the exterior velocity field given by Eq.~\eqref{OutsideSolutionAxi} by $8\pi\eta^{(i)}$ rather than by $8\pi\eta^{(e)}$.

Having expressed the velocity field on both sides of the drop in terms of harmonics based at the origin, we next determine the unknown series coefficients $\{A_n^\parallel , B_n^\parallel \}$ and $\{a_n^\parallel, b_n^\parallel\}$. 
By applying the boundary conditions prescribed at the surface of the drop, given by Eq.~\eqref{boundaryCond1} and~\eqref{boundaryCond2clean} for a clean drop, and by Eqs.~\eqref{boundaryCond1} and~\eqref{boundaryCond2surfactant} for a surfactant-covered drop, and using the fact that $\boldNabla \varphi_n$ and $\R \varphi_n$ form a set of orthogonal vector harmonics, we obtain a system of linear equations. Its solution yields the expressions of the series coefficients  associated with the solution of the flow field inside and outside the drop.
Further details of derivation are shifted to Appendix~\ref{appendix:1}.
For a clean drop, the coefficients are given by 
\begin{subequations}\label{Coeffs_Parallel}
	\begin{align}
		A_n^\parallel  &= \left( \Lambda-1 + \left( \frac{2n+1}{2n+3} - \Lambda \right)R^{2} \right) R^{n-1} \, , \\
		B_n^\parallel  &= \frac{n+1}{2}  \left( \frac{2n+1}{2n-1} - \Lambda + \left(\Lambda-1\right)R^{2} \right) R^{n-1} \, , \\
		a_n^\parallel &= \Lambda \left( 1 - R^{2} \right) R^{n-1} \, , \\
		b_n^\parallel &= \frac{\Lambda n}{2} \left( 1 - R^{2} \right) R^{n-1} \, , 
	\end{align}
\end{subequations}
where we have defined for convenience the dimensionless number $\Lambda = \lambda/(1+\lambda)$ with $\lambda = \eta^{(i)} / \eta^{(e)}$ denoting the viscosity contrast.
Accordingly, $\Lambda$ vanishes in the rigid-cavity limit (\textit{e.g.}\ water drop in extremely viscous oil) and approaches one for drops with a large viscosity compared to the external medium (\textit{e.g.}\ water drop in air).

For a surfactant-covered drop, it follows from Eq.~\eqref{gradSVs2} that the surface velocity vanishes in the axisymmetric case.
Accordingly, the solution of the axisymmetric flow problem for a Stokeslet acting inside a viscous drop covered with a non-diffusing, insoluble, and incompressible layer of surfactant is identical to that inside a rigid spherical cavity $(\Lambda=0)$. 
Specifically,
\begin{subequations}
	\begin{align}
		A_n^\parallel  &= \left( \frac{2n+1}{2n+3} \, R^{2} -1 \right) R^{n-1} \, , \\
		B_n^\parallel  &= \frac{n+1}{2}  \left( \frac{2n+1}{2n-1} -R^{2} \right) R^{n-1} \, , \\
		a_n^\parallel &= b_n^\parallel = 0\, .
	\end{align}
\end{subequations}
It is worth mentioning that analogous behavior has been found for a Stokeslet acting near a planar interface covered with surfactant~\cite{blawzdziewicz99} and for a Stokeslet acting outside a surfactant-covered drop~\cite{shaik17}.

\subsection{Transverse Stokeslet}

We proceed in an analogous way as in the axisymmetric case and express the velocity field on both sides of the drop in terms of harmonics based at~$\vect{x}_1$.
As demonstrated in detail in Ref.~\cite{hoell19creeping}, the free-space Stokeslet solution for a transverse point force $\vect{F} = F^\perp \vect{e}$ acting at the position~$\vect{x}_2$ can be written via Legendre expansion as an infinite series as
\begin{equation}
 8\pi\eta^{(i)}\vStok = F^\perp \infSumOne \left( \beta_n  \boldNabla\psi_{n-1}
                 -\frac{2 R^n}{n+1} \, \boldsymbol{\gamma}_{n-1} 
                 + \boldsymbol{\tau}_n \right) \, , \label{StokesletAsymm}
\end{equation}
where
\begin{subequations}
	\begin{align}
		\beta_n &= \left( \frac{n-2}{n(2n-1)} \, r^2 - \frac{n R^{2}}{(n+2)(2n+3)} \right) R^{n-1} \, , \\
		\boldsymbol{\tau}_n  &= - \frac{2(n+1) R^{n-1}}{n(2n-1)} \, \vect{r}  \psi_{n-1},
	\end{align}
\end{subequations}
and where we have defined the harmonics	$\psi_n = (\vect{e} \cdot\boldNabla)\varphi_{n}$ and $\boldsymbol{\gamma}_n = \vect{t}  \times \boldNabla \varphi_{n}$, with the unit vector $\vect{t} = \vect{e} \times \vect{d}$.
By construction, $\psi_n = \boldsymbol{\gamma}_n \cdot \vecD$. In contrast to the simple axisymmetric case for which only two orthogonal vector harmonics are needed as basis function for the expansion of the flow field, the transverse situation requires three vector harmonics that we chose here for convenience to be~$\boldNabla \psi_n$, $\R \psi_n$, and~$\boldsymbol{\gamma}_n$.

In addition, the image solution inside the drop can likewise be obtained using Lamb's general solution and be expressed in terms of \emph{exterior} harmonics as~\cite{hoell19creeping}
\begin{equation}
		8\pi\eta^{(i)}\vect{v}^{*} = F^\perp \infSumOne 
		\left( A_n^\perp \vect{Q}_n^{(1)} + B_n^\perp \vect{Q}_n^{(2)} + C_n^\perp \vect{Q}_n^{(3)} \right) \, , \label{ImageSolutionAsymm}
\end{equation}
where we have defined the vector functions
\begin{subequations}\label{Qn}
	\begin{align}
		\vect{Q}_n^{(1)} &=  \left( \frac{n+3}{2n} \, r^{2} \boldNabla\psi_{n-1}  
				+ \frac{(n+1)(2n+3)}{2n} \, \R \psi_{n-1} \right) r^{2n+1} \, , \\
		\vect{Q}_n^{(2)} &= \frac{1}{n} \bigg( r^2 \, \boldNabla\psi_{n-1} 
				+ (2n+1) \, \R \psi_{n-1} \bigg) r^{2n-1} \, , \\
		\vect{Q}_n^{(3)} &= \left( \boldsymbol{\gamma}_{n-1} 
				+ \frac{2n-1}{r^2} \, (\vect{t} \times\R) \varphi_{n-1} \right) r^{2n-1} \, . 
	\end{align}
\end{subequations}

Finally, the solution of the flow problem outside the spherical drop can be expressed in terms of \emph{exterior} harmonics as~\cite{hoell19creeping}
\begin{align}
		8\pi\eta^{(i)} \vect{v}^{(e)} &= F^\perp \infSumOne  
			\Bigg( a_n^\perp \left( \frac{n-2}{2(n+1)} \, r^2 \boldNabla \psi_{n-1} - \vect{r} \psi_{n-1}  \right) \notag \\
			&\quad-\frac{b_n^\perp}{n+1} \, \boldNabla \psi_{n-1}
			+ c_n^\perp \boldsymbol{\gamma}_{n-1} \Bigg) \, , \label{outsideSolution_finalized}
	\end{align}
where, again, we have chosen, for the sake of convenience, to rescale the exterior flow field by $8\pi\eta^{(i)}$ rather than by $8\pi\eta^{(e)}$.

For a clean drop, solving for the series coefficients $\{ A_n^\perp, B_n^\perp, C_n^\perp \}$ and $\{ a_n^\perp, b_n^\perp, c_n^\perp \}$ associated with the flow fields inside and outside the drop, respectively, yields {\allowdisplaybreaks
\begin{subequations}\label{Coeffs_Perp_Clean}
	\begin{align}
		A_n^\perp &= \left(\Lambda - 1 + \frac{n+3}{n+1} \left( \frac{2n+1}{2n+3} - \Lambda \right) R^2 \right) R^{n-1} \, , \\
		B_n^\perp &= \bigg( (n+1) k_n 
		- \frac{n+3}{2} \left( 1-\Lambda \right) R^{2} \bigg) R^{n-1}  \, , \\
		C_n^\perp &= \frac{2n (1-2\Lambda) R^{n-2}}{(n-2) \left(3\Lambda-n\right)} \, , \\
		a_n^\perp &= \frac{\Lambda}{n} \left( (n+3)R^2-n-1 \right) R^{n-1} \, , \\
		b_n^\perp &= \Lambda  \bigg(  \left( g_{n+1}+ \frac{n+3}{2} \right) R^2 - \frac{n+1}{2} \bigg) R^{n-1} \, , \\
		c_n^\perp &= \Lambda g_n \, R^{n} \, , 
	\end{align}
\end{subequations}
}where we have defined
\begin{subequations}
	\begin{align}
		k_n &= \frac{2(1-2\Lambda)}{(n-1) \left(3\Lambda-n-1\right)} + \frac{2n+1}{2(2n-1)} - \frac{\Lambda}{2} \, , \\
		g_n &= \frac{2 (2n+1)}{(n+1) \left( 3\Lambda - n - 2 \right)} \, .
	\end{align}
\end{subequations}

For a surfactant-covered drop, the corresponding coefficients are given by
\begin{subequations}\label{Coeffs_Perp_Surfactant}
	\begin{align}
		A_n^\perp &= \left( \frac{\left( n+3 \right) \left(2n+1\right)}{\left( n+1 \right) \left( 2n+3 \right)} \, R^2 - 1 \right) R^{n-1} \, , \\
		B_n^\perp &= \frac{1}{2} \left( \frac{\left( n+1 \right) j_n}{\left(n-1\right) \left(2n-1\right) h_n}  - \left(n+3\right) R^2 \right) R^{n-1} \, , \\
		C_n^\perp &= \frac{2n \left( \lambda+3w-1-wn \right) R^{n-2}}{\left(n-2\right) \left( wn^2 + \left(1+\lambda-3w\right)n-3\lambda \right)} \, , \\
		a_n^\perp &= 0 \, , \\
		b_n^\perp &= -\frac{2 \lambda\left(2n+3\right) R^{n+1}}{(n+2) \left( wn^2 + \left(1+\lambda+3w\right)n+3 \right)} \, , \\
		c_n^\perp &= b_{n-1}^\perp \, , 
	\end{align}
\end{subequations}
where we have defined
\begin{equation}
	w = \frac{\eta_\mathrm{S}}{\eta^{(e)}}
\end{equation}
as an inverse length parameter.
In addition,
\begin{subequations}
	\begin{align}
		j_n &= 2w n^4 + \left(2+2\lambda-3w\right)n^3 + \left(1-5\lambda-12w\right)n^2 \notag \\
		&\quad+ \left( 9\lambda+23w-10 \right)n + 3-2\lambda-6w \, , \\
		h_n &= wn^2 + \left( 1+\lambda-w \right)n + 1-2\lambda-2w \, .
	\end{align}
\end{subequations}
For further details of derivation, we refer to Appendix~\ref{appendix:1}.
Notably, the series coefficients $A_n^\perp$ and $a_n^\perp$ for a surfactant-covered drop are equal to those for a rigid spherical cavity $(\Lambda=0)$.
We note that the rigid cavity limit is recovered for all the other series coefficients by taking the limits $\lambda \to 0$ (or alternatively $\Lambda\to 0$).

In the limit $w \to \infty$, the fluid flow outside the cavity is described by the only non-vanishing coefficient $c_1^\perp = -\lambda R$.

\subsection{Solution for a freely moving drop}

So far, we have assumed that the fluid velocity normal to the interface of the drop vanishes that the drop remains at rest.
This implies that in general an external force has to be exerted on the drop to maintain it at its present location.
The additionally applied force is equal in magnitude but different in sign when compared to the hydrodynamic force exerted by the Stokeslets on the stationary drop.
Accordingly, the solution of the flow problem for a freely moving drop can be obtained by accounting for the Stokeslet solution derived above and adding a flow field induced by a drop subject to an external force that just balances the force applied previously to maintain the drop in position.

For a Stokeslet acting inside a stationary drop, the hydrodynamic force against the flow of the outside fluid is obtained by integrating the traction vector on the \emph{outer} surface of the drop as~\cite{leal07book}
\begin{equation}
	\vect{F}_\mathrm{Drop}^\mathrm{S} = \int_0^{2\pi} \int_0^\pi \vect{T}^{(e)} \sin\theta \, \Intd \theta \, \Intd \phi \, ,  
\end{equation} 
which after calculation leads to
\begin{equation}
	\vect{F}_\mathrm{Drop}^\mathrm{S} = \lambda^{-1} \left( 
	a_1^\parallel F^\parallel \, \vect{d}
	- \frac{a_1^\perp}{4} \, F^\perp \, \vect{e} \right) . 
	\label{HydroForceStokeslet} 
\end{equation}
This force is necessary to be imposed on the surface of the drop to maintain it in position, which ensures the surface condition in Eq.~(\ref{VanishingRadial}). For a rigid cavity the flow field outside the cavity vanishes, $a_1^\parallel = a_1^\perp = 0$, and thus the cavity does not experience any force.
Upon substitution of the two series coefficients~$a_1^\parallel$ and~$a_1^\perp$, we obtain for a clean drop
\begin{equation}
	\vect{F}_\mathrm{Drop}^\mathrm{S} = 
	 \frac{1-\Lambda}{2} \left( \left( 1-R^2 \right) F^\parallel \, \vect{d} 
	 +  \left( 1-2R^2 \right) F^\perp \, \vect{e} \right) .
\end{equation}

The resulting translational velocity can then be obtained as $\vect{V}_\mathrm{Drop}^\mathrm{S} = \mu \vect{F}_\mathrm{Drop}^\mathrm{S}$, with $\mu = 1/ \left( 2\pi \left(2+\Lambda\right) \eta^{(e)} \right)$ denoting the translational hydrodynamic mobility of a clean drop. We find
\begin{equation}
	\vect{V}_\mathrm{Drop}^\mathrm{S} = 
	\frac{\left( 1-\Lambda \right) \Big( \left( 1-R^2 \right) F^\parallel \, \vect{d} 
	+  \left( 1-2R^2 \right) F^\perp \, \vect{e} \Big)}{4\pi \left(2+\Lambda\right) \eta^{(e)}}	\, . \label{VDrop}
\end{equation}

The axisymmetric flow field induced by a drop translating with a constant velocity~$V \vect{d}$ is known as the \textit{Hadamard-Rybczynski} solution and can be found in classic fluid mechanics textbooks.
It is given in the frame of the drop by~\cite[ch.~7, p.~482]{leal07book}
\begin{subequations}\label{hadamardOut}
	\begin{align}
		v_r^{(e)} &= -V\left(1 - \frac{2+\Lambda}{2r} + \frac{\Lambda}{2r^3}\right) \cos\theta \, , \\
		v_\theta^{(e)} &= V\left( -1 + \frac{2+\Lambda}{4r} + \frac{\Lambda}{4r^3} \right) \sin\theta \, , 
	\end{align}
\end{subequations}
for the outer fluid, and by 
{\allowdisplaybreaks
\begin{subequations}\label{hadamardIns}
	\begin{align}
		v_r^{(i)} &= \frac{V}{2} \left(1-\Lambda\right) \left(1-r^2\right) \cos\theta \, , \\
		v_\theta^{(i)} &= \frac{V}{2} \left(1-\Lambda\right) \left(1-2r^2\right) \sin\theta \, ,
	\end{align}
\end{subequations}
}for the inner fluid. 
Consequently, the total flow field induced by a Stokeslet acting inside a freely movable drop is obtained by superimposing the flow field resulting from a Stokeslet acting inside a stationary drop and the flow field induced by a drop translated with a constant velocity~$-\vect{V}_\mathrm{Drop}^\mathrm{S}$. That is, we impose a flow field that is in principle resulting from a force $-\vect{F}_\mathrm{Drop}^\mathrm{S}$ added to cancel the force $\vect{F}_\mathrm{Drop}^\mathrm{S}$ that we had effectively imposed before to keep the drop in position and thus to satisfy Eq.~(\ref{VanishingRadial}). 


\begin{figure}
	\includegraphics[scale=0.6]{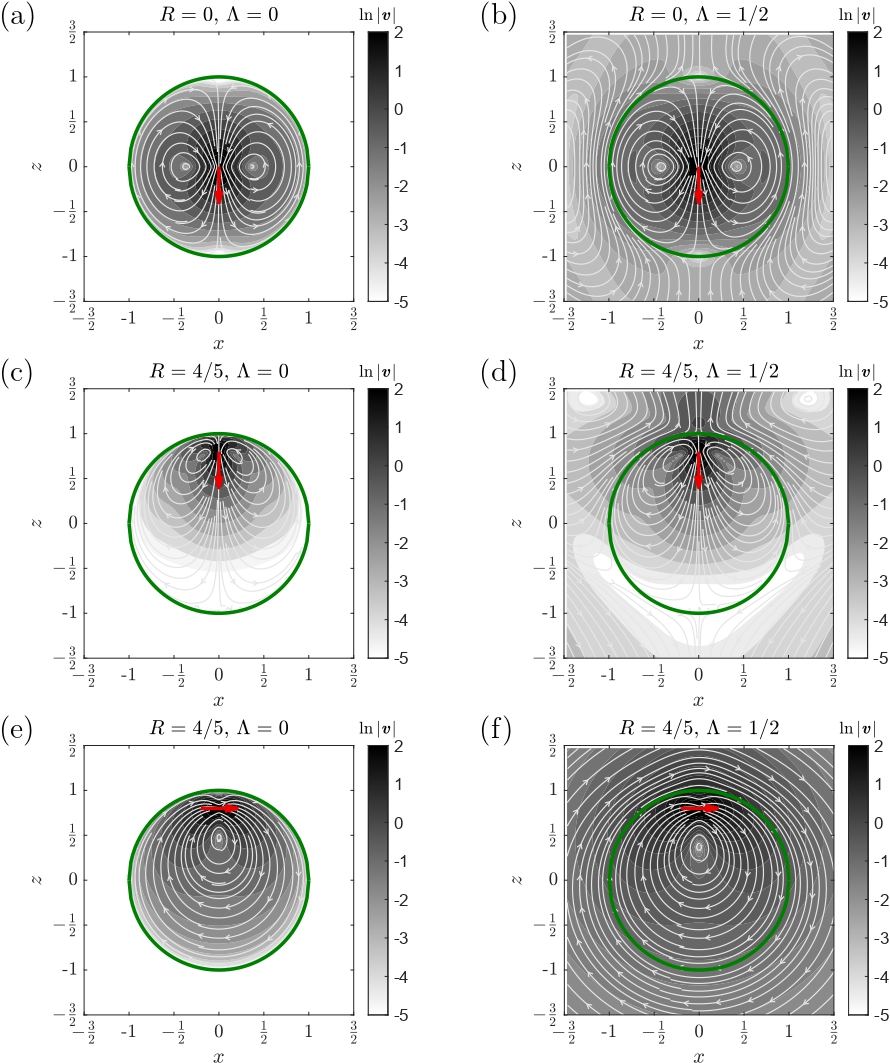}
	\caption{(Color online) Streamlines and contour plots of the flow field induced by an axisymmetric [(a) -- (d)] and transverse Stokeslet [(e) and~(f)] inside a clean freely moving drop for different values of~$R$ and~$\Lambda$.
		The Stokeslet singularity is represented by a red one-headed arrow.
		In the left column, $\Lambda=0$ corresponds to a rigid spherical cavity, while the right column of $\Lambda=1/2$ allows flow fields to be induced in the outer fluid by the presence of a point force inside the drop. 
		The velocity magnitude is scaled by $1/ \left(8\pi\eta^{(i)}\right)$.
		To indicate the magnitude of the flow field, shading is used on a logarithmic scale.}
	\label{Mono}
\end{figure}
\begin{figure}
	\includegraphics[scale=0.6]{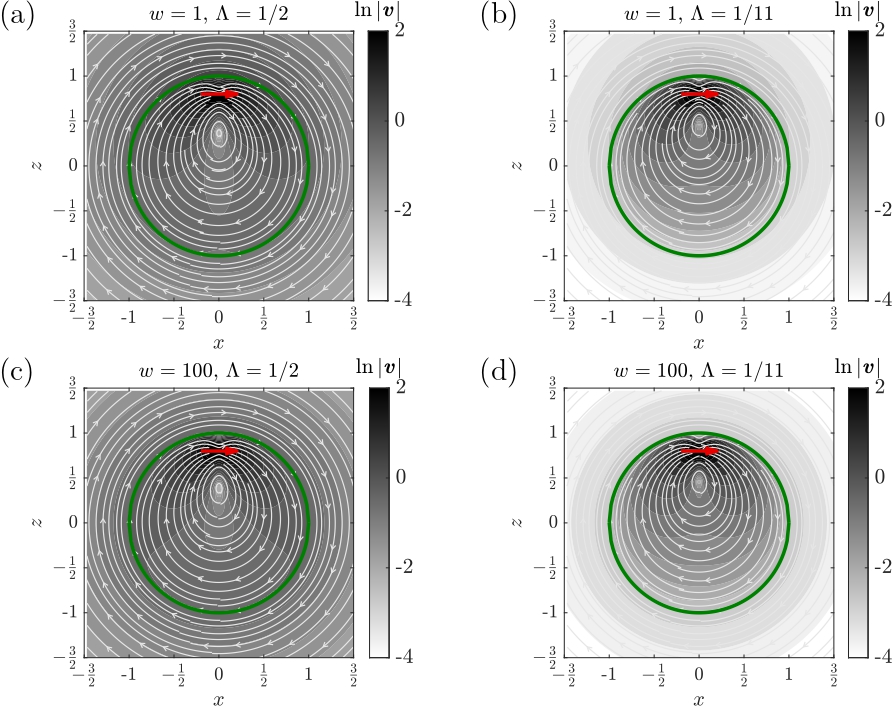}
	\caption{(Color online) Streamlines and contour plots of the flow field induced by a transverse Stokeslet inside a surfactant-covered drop for $R=4/5$ and two different values for~$\Lambda$ and~$w$.
		The transverse Stokeslet is represented by a red one-headed arrow.
		The inner flow field resembles that of a clean drop whereas the outer flow field consists of circular streamlines.
		Here, the velocity magnitude is scaled by $1/ \left(8\pi\eta^{(i)}\right)$.}
	\label{Mono-Surfactant}
\end{figure}

For a surfactant-covered drop, we have shown that $a_1^\parallel = a_1^\perp = 0$.
Thus for a rigid cavity and a surfactant-covered drop the total net force transmitted to the drop vanishes.
This similarity can be motivated as follows. 
Any flow past a spherical surface, irrespective of boundary conditions on the surface, can be decomposed into a surface solenoidal and a surface irrotational flow field on the spherical surface~\cite{shaik17}. 
The surface irrotational flow field is torque-free and it exerts a force and a stresslet on the particle, whereas the surface solenoidal flow field is force-free and stresslet-free and it exerts a torque on the particle. 
For a viscous drop (both clean and surfactant-covered), the surface solenoidal flow field is additionally torque-free~\cite{chan77}. 
For a drop covered with an incompressible surfactant of zero surface diffusivity, the surface irrotational flow field is the same as that of a rigid spherical cavity~\cite{shaik17}. 
For this reason, the force and stresslet experienced by a surfactant-covered drop are the same as those experienced by a rigid spherical cavity, regardless of the specific value of the viscosity contrast $\lambda$ and surface to external bulk viscosity ratio $w$.

The resulting flow fields can now be computed for an arbitrary position and viscosity ratio. As an illustration, in Fig.~\ref{Mono}, we draw on the left-hand side the streamlines and the magnitude of the flow field created by a point force inside a stiff spherical cavity (for $\Lambda=0$), which coincides with the well-known image solution~\cite{Maul1994}. On the right-hand side, we depict the case of $\Lambda=1/2$ for a freely moving drop in the absence of the surfactant. 
Here, the flow inside the drop induces motion of the exterior fluid. 
The magnitude of the flow velocity fields is shown on a logarithmic scale.
In particular, the case of stiff confinement (left column) leads to a faster decay of the velocity magnitude due to an increased dissipation at the boundary. For the radially-oriented Stokeslet, the patterns retain rotational symmetry about the Stokeslet direction. 
Accordingly, the flow field inside the drop consists of toroidal eddies owing to the axisymmetric nature of the flow~\cite{moffatt64}.
In contrast to that, a single vortex is created inside the drop for the transverse point force.

In Fig.~\ref{Mono-Surfactant}, we include the effect of the surfactant by examining two non-zero values of~$\Lambda$ and~$w$.
The non-vanishing surfactant shear viscosity does not change qualitatively the shape of the streamlines inside the drop.
However, the outer fluid shows concentric circular streamlines similar to those resulting from the uniform rotation of a rigid body.
For a fixed viscosity contrast, we observe a weak dependence of the velocity magnitude on the parameter~$w$, whereas the topology and structure of the flow field remain nearly invariant in the investigated parameter regime

Having derived the image solution for a point-force singularity acting inside a spherical viscous drop, we next make use of this solution to derive the corresponding image for a force dipole singularity.

\section{Dipole singularity}\label{sec:Dipo}

In the following, we denote by~$\vect{q} := \vect{F} / \left|\vect{F}\right| $ the unit vector pointing along the direction of the force.
Additionally, we define the Green's function associated with the
$\vect{q}$-directed Stokeslet acting at the position~$\vect{x}_2$ of an unbounded fluid medium as
\begin{equation}
	\vect{G}(\vect{q}) = 8\pi\eta^{(i)} \, \boldsymbol{\G} (\X-\X_2) \cdot \vect{q} \, .
	\label{Gq}
\end{equation}

In the far-field limit, the force monopole decays with inverse distance from the singularity position. 
For an arbitrary orientation of the Stokeslet, the unit vector~$\vect{q}$ can be projected along the axisymmetric and transverse directions as
\begin{equation}
	\vect{q} = \sin\delta \, \vect{d} + \cos\delta \, \vect{e} \, .  \label{q-Projection}  
\end{equation}

The flow field induced by force- and torque-free swimming microorganisms can be written as a multipole expansion of the solution of the Stokes equations~\cite{spagnolie12}.
To leading order, this flow field appears as induced by a force dipole, which exhibits a decay with inverse distance squared and thus faster than flows induced by force monopoles. 
Higher-order singularities associated with Stokes flows can be obtained by differentiations of the Stokeslet solution.

We define the free-space flow field caused by a force dipole as
\begin{equation}
	\vect{G}_\mathrm{D} (\vect{q}, \vect{p}) = \left( \vect{p} \cdot \boldsymbol{\nabla} \right)
	\boldsymbol{G}(\vect{q}) \, ,
\end{equation}
where $\vect{p}$ is a unit vector along which the gradient operator is exerted.
In an unbounded fluid medium, \textit{i.e.}, for an infinitely large radius of the drop, the self-generated flow induced by an active force-dipole model microswimmer oriented along the direction of the unit vector~$\vect{q}$ is expressed as $\vect{v}_\mathrm{D} = -\alpha \,  \vect{G}_\mathrm{D} (\vect{q}, \vect{q} )$.
Accordingly,
\begin{equation}
		\vect{v}_\mathrm{D}
		= -\alpha \left( \vect{q} \cdot \boldNabla  \right)
		\left( \sin \delta \, \vect{G}(\vect{d}) + \cos \delta \, \vect{G}(\vect{e}) \right) \, ,
		\label{vD-bulk}
\end{equation}
where~$\alpha$ sets the strength of the force dipole.
Then, for a general orientation, the force dipole can be written as a linear combination of axisymmetric and transverse force dipole singularities as
{\allowdisplaybreaks
\begin{align}
	\vect{G}_\mathrm{D} (\vect{q}, \vect{q} ) 
	&= \vect{G}_\mathrm{D} (\vect{d}, \vect{d} ) \, \sin^2\delta
	+ \vect{G}_\mathrm{D} (\vect{e}, \vect{e} ) \, \cos^2\delta \notag  \\[2pt]
	&\quad+ \underbrace{ \tfrac{1}{2} \big( \vect{G}_\mathrm{D} (\vect{e}, \vect{d} ) + \vect{G}_\mathrm{D} (\vect{d}, \vect{e} ) \big) }_\text{$\vect{G}_\mathrm{SS} (\vect{e}, \vect{d} )$} \, \sin \left(2\delta\right) \, ,  \label{GD-General}
\end{align}
}where~$\vect{G}_\mathrm{SS} (\vect{e}, \vect{d} ) = \vect{G}_\mathrm{SS} (\vect{d}, \vect{e} )$ stands for the symmetric part of the Green's function associated with the force dipole, which is commonly termed the stresslet~\cite{lopez14}.

We now summarize the main mathematical operations required for the calculation of each of the image flow fields resulting for Eq.~\eqref{GD-General}.
Denoting by $\mathbb{I} \{ \vect{v}  \}$ the image solution for a given flow field~$\vect{v}$, it can be shown that~\cite{shaik17, fuentes88, fuentes89}
\begin{subequations}
	\begin{align}
		\mathbb{I} \{  \vect{G}_\mathrm{D} (\vect{d}, \vect{d} ) \} &= 
		-\left( \vect{d} \cdot \boldNabla_2 \right) \mathbb{I} \{ \vect{G}(\vect{d} ) \} \, , \label{Im1} \\
		\mathbb{I} \{  \vect{G}_\mathrm{D} (\vect{e}, \vect{e} ) \} &= 
		-\left( \vect{e} \cdot \boldNabla_2 \right) \mathbb{I} \{ \vect{G}(\vect{e} ) \}
		+ R^{-1} \, \mathbb{I} \{ \vect{G}(\vect{d} ) \} \, , \label{Im2} \\
		\mathbb{I} \{  \vect{G}_\mathrm{D} (\vect{e}, \vect{d}) \} &= 
		-\left( \vect{d} \cdot \boldNabla_2 \right) \mathbb{I} \{ \vect{G}(\vect{e} ) \} \, , \label{Im3} \\
		\mathbb{I} \{ \vect{G}_\mathrm{D} (\vect{d}, \vect{e}) \} &= 
		-\left( \vect{e} \cdot \boldNabla_2 \right) \mathbb{I} \{ \vect{G}(\vect{d} ) \}
		- R^{-1} \, \mathbb{I} \{ \vect{G}(\vect{e} ) \} \, . 
		\label{Im4}
	\end{align}
\end{subequations}
Here, we have made use of the relations $\left( \vect{e} \cdot \boldsymbol{\nabla}_2 \right) R = 0$, $\left( \vect{e} \cdot \boldsymbol{\nabla}_2 \right) \vect{d} = - (1/R) \vect{e}$, and $\left( \vect{e} \cdot \boldsymbol{\nabla}_2 \right) \vect{e} = (1/R) \vect{d}$.

By noting that $\vect{d} \cdot \boldNabla_2 = - \partial / \partial R $, it follows from Eqs.~\eqref{ImageSolutionAxi} and \eqref{Im1} that the image solution for the axisymmetric force dipole can be expressed as
\begin{equation}
	\mathbb{I} \{  \vect{G}_\mathrm{D} (\vect{d}, \vect{d} ) \}
	= \sum_{n=1}^{\infty} \left( 
		\frac{\partial A_n^\parallel}{\partial R} \, \vect{S}_n^{(1)}
		+ \frac{\partial B_n^\parallel}{\partial R} \, \vect{S}_n^{(2)}\right) \, , \label{dGradd} 
\end{equation}
where the vector functions~$\vect{S}_n^{(j)}$ ($j\in\{1,2\}$) involve the harmonics~$\boldsymbol{\nabla} \varphi_n$ and~$\vect{r} \varphi_n$, and have previously been defined by Eqs.~\eqref{Sn}.

In addition, it follows from Eqs.~\eqref{ImageSolutionAsymm} and~\eqref{Im3} that
\begin{equation}
		\mathbb{I} \{  \vect{G}_\mathrm{D} (\vect{e}, \vect{d}) \} = 
		\infSumOne 
		\left( \frac{\partial A_n^\perp}{\partial R} \, \vect{Q}_n^{(1)} 
		+ \frac{\partial B_n^\perp}{\partial R} \, \vect{Q}_n^{(2)}
		+ \frac{\partial C_n^\perp}{\partial R} \, \vect{Q}_n^{(3)} \right) ,
		\label{dGrade}
\end{equation}
where the vector functions~$\vect{Q}_n^{(j)}$ ($j\in\{1,2,3\}$) involve the harmonics~$\boldsymbol{\nabla} \psi_{n-1}, \vect{r} \psi_{n-1}$, and $\boldsymbol{\gamma}_{n-1}$, see the definitions in Eqs.~\eqref{Qn}.

Involving the relation
\begin{equation}
		\left( \vect{e} \cdot \boldsymbol{\nabla}_2 \right) \varphi_n 
		= -R^{-1} \left( \vect{e} \cdot \boldsymbol{\nabla} \right) \varphi_{n-1}
		= -R^{-1} \, \psi_{n-1} \, ,
\end{equation}
we readily obtain
\begin{equation}
	\left(\vect{e} \cdot \boldsymbol{\nabla}_2 \right)
	\mathbb{I} \{ \vect{G}(\vect{d} ) \} 
    = -\frac{1}{R} \, \sum_{n=1}^{\infty} n \left( 
    A_n^\parallel \vect{Q}_n^{(1)}  
    + B_n^\parallel \vect{Q}_n^{(2)}  \right)  \, . \label{eGradd}
\end{equation}

Combining results, the image of the stresslet field can be cast in the final form
\begin{equation}
		\mathbb{I} \{ \vect{G}_\mathrm{SS}(\vect{d}, \vect{e} ) \} = 
		\sum_{n=1}^{\infty}
		\left( 
		\hat{A}_n \vect{Q}_n^{(1)} 
		+ \hat{B}_n \vect{Q}_n^{(2)}
		+\hat{C}_n \vect{Q}_n^{(3)}
		\right) \, ,
\end{equation}
where the series coefficients are given by
{\allowdisplaybreaks
\begin{subequations}
	\begin{align}
		\hat{A}_n &= \frac{1}{2} \left( \frac{n}{R} \, A_n^\parallel + R \, \frac{\partial}{\partial R}	\left( \frac{A_n^\perp}{R} \right) \right) \, , \\
		\hat{B}_n &= \frac{1}{2} \left( \frac{n}{R} \, B_n^\parallel + R \, \frac{\partial}{\partial R} \left( \frac{B_n^\perp}{R} \right) \right) \, , \\
		\hat{C}_n &= \frac{R}{2}\frac{\partial}{\partial R} \left( \frac{C_n^\perp}{R} \right) \, .	
	\end{align}
\end{subequations}}

Next, by making use of the relation
\begin{subequations}
	\begin{align}
	\left( \vect{e} \cdot \boldsymbol{\nabla}_2 \right) \psi_{n-1}
		&= R^{-1} \big( \left( \vect{d} \cdot \boldsymbol{\nabla} \right) \varphi_{n-1} 
		- \left( \vect{e} \cdot \boldsymbol{\nabla}\right) \psi_{n-2} \big) \notag \\
		&= R^{-1} \left( n \varphi_n - \xi_{n-2} \right) \, , 
	\end{align}
\end{subequations}
together with $\xi_n := \left( \vect{e} \cdot \boldsymbol{\nabla} \right) \psi_n$, we readily obtain 
\begin{align}
	\left(\vect{e} \cdot \boldsymbol{\nabla}_2 \right)
		\mathbb{I} \{ \vect{G}(\vect{e} ) \} 
		= \infSumOne
		\frac{1}{R}
		\left( A_n^\perp \vect{W}_n^{(1)} + B_n^\perp \vect{W}_n^{(2)} + C_n^\perp \vect{W}_n^{(3)} \right) \, , 
\end{align}
where we have defined the vector functions
\begin{subequations}
	\begin{align}
		\vect{W}_n^{(1)} &=  \vect{S}_n^{(1)} - \frac{1}{2n} \bigg( (n+3) \, r^{2}   
		 \boldNabla \xi_{n-2}
				+ \rho_n \, \R \, \xi_{n-2} \bigg) r^{2n+1}  \, , \notag \\
		\vect{W}_n^{(2)} &=  \vect{S}_n^{(2)} - \frac{1}{n} \bigg( r^2 \, \boldNabla \xi_{n-2} 
				+ (2n+1) \, \R \, \xi_{n-2} \bigg) r^{2n-1} \, , \notag \\
		\vect{W}_n^{(3)} &= -\left( \vect{t} \times \boldNabla \psi_{n-2} + \frac{2n-1}{r^2} \, (\vect{t} \times\R) \psi_{n-2} \right) r^{2n-1} \, , \notag
	\end{align}
\end{subequations}
together with $\rho_n = (n+1)(2n+3)$.

Having derived the image flow field for a force dipole singularity acting inside a stationary drop, we next determine the external force that is needed to maintain the drop in position, which corresponds to the condition of vanishing normal velocity at the interface imposed by Eq.~(\ref{VanishingRadial}).

\begin{figure}
	\includegraphics[scale=0.6]{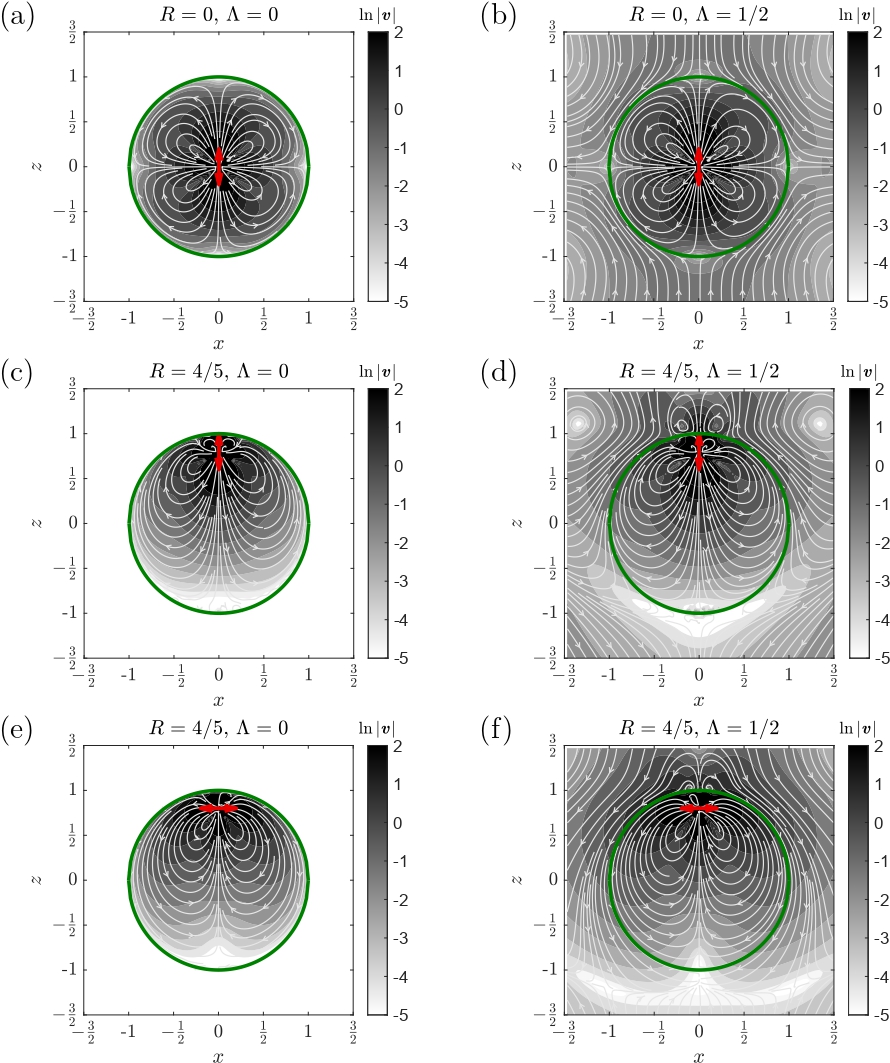}
	\caption{(Color online) Streamlines and contour plots of the flow field induced by an axisymmetric (a-d) and transverse force dipole (e-f) inside a clean drop for different positions~$R$, orientations, and values of $\Lambda$. Similarly to the case of a point force, see Fig.~\ref{Mono}, for the effectively stiff spherical cavity (left column), we observe by construction a quick decay of the flow field towards the boundary of the drop. In the case of equal viscosity inside and outside the drop (right column), we find additional recirculation zones appearing close to the surface of the freely moving drop.}
	\label{Dipo}
\end{figure}
\begin{figure}
	\includegraphics[scale=0.6]{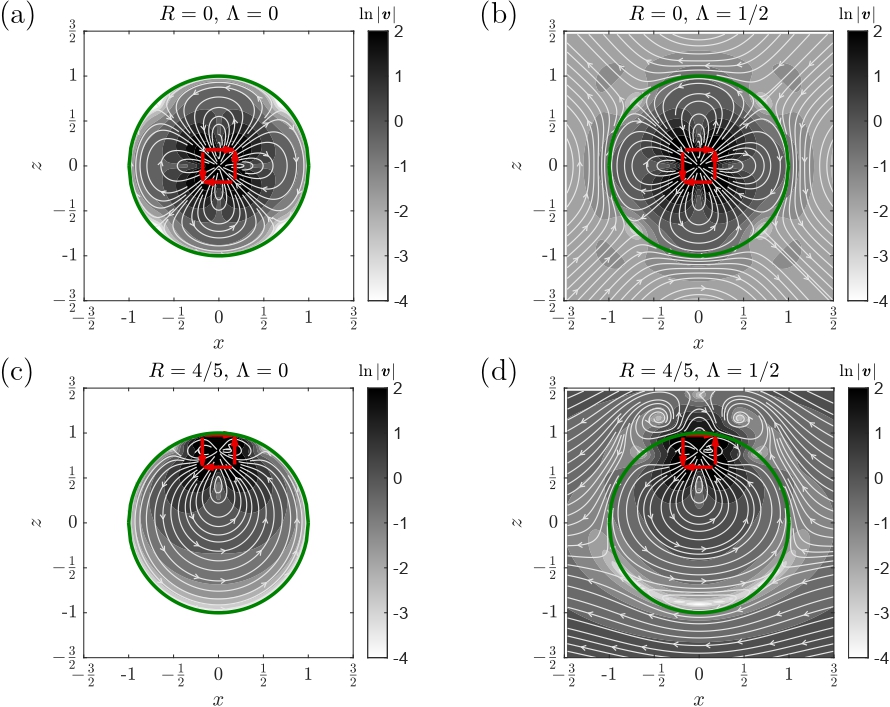}
	\caption{(Color online) Streamlines and contour plots of the flow field induced inside a clean drop by a stresslet placed at the origin [(a) and~(b)] or off-center [(c) and (b)] for two different values of the viscosity contrast, corresponding to an effectively stiff boundary ($\Lambda=0$) and to an equal viscosity of the inner and outer fluids for a freely moving drop.
		The flow far away from the drop, retains the same geometric signature as the generating stresslet.}
	\label{Stresslet}
\end{figure}
\begin{figure}
	\includegraphics[scale=0.6]{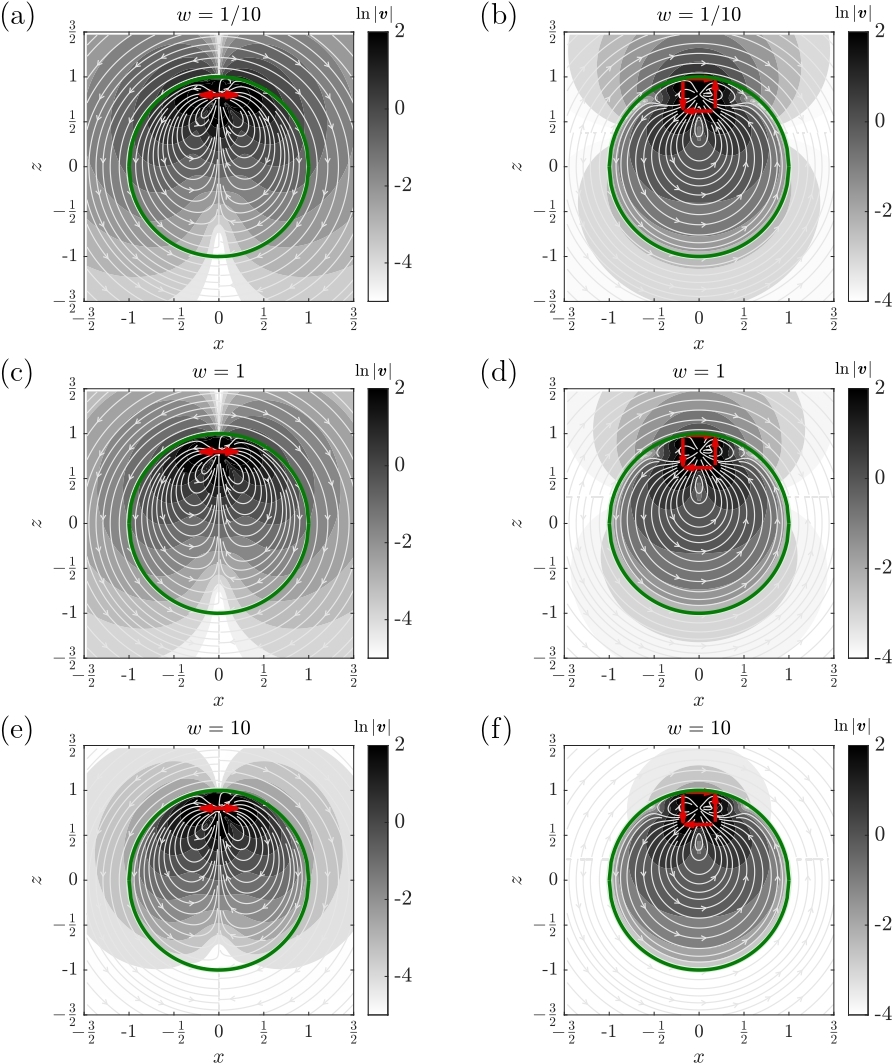}
	\caption{(Color online) Streamlines and contour plots of the flow field induced by a force dipole [(a), (c), and~(e)] and a stresslet [(b), (d), and~(f)] for $R=4/5$, $\Lambda=1/2$, and three different values of~$w$ when the surface of the drop contains a surfactant. 
		The structure of the streamlines is qualitatively different from that of a clean drop.}
	\label{Dipo-Stresslet-Surfactant}
\end{figure}

The hydrodynamic force against the outside fluid flow in the presence of the  force dipole can again be obtained by integrating the hydrodynamic traction vector on the \emph{outer} surface as~\cite{leal07book}
\begin{equation}
	\vect{F}_\mathrm{Drop}^\mathrm{D} = \int_0^{2\pi} \int_0^\pi \vect{T}_\mathrm{D}^{(e)} \sin\theta  \, \Intd \theta \, \Intd \phi \, ,  
\end{equation} 
which leads to
\begin{equation}
	\vect{F}_\mathrm{Drop}^\mathrm{D} = -\alpha \pi \eta^{(e)}
	\left( 2f_\parallel  \vect{d} + f_\perp \vect{e} \right) \, ,
	\label{HydroForceDipole}
\end{equation}
together with the definitions
\begin{subequations}
	\begin{align}
		f_\parallel &= \frac{2a_1^\parallel+a_1^\perp}{R} \, \cos^2\delta - 2 \, \frac{\partial a_1^\parallel}{\partial R} \, \sin^2\delta \, , \\
		f_\perp &= \left( \frac{2a_1^\parallel+a_1^\perp}{R} - \frac{\partial a_1^\perp}{\partial R} \right) \sin (2\delta) \, .
	\end{align}
\end{subequations}
Upon substitution of the series coefficients, we readily obtain for a clean viscous drop 
\begin{equation}
	\vect{F}_\mathrm{Drop}^\mathrm{D} = -2\alpha\pi \eta^{(e)} R \Lambda 
	\Big( \left( 3-\cos(2\delta) \right) \vect{d} - 3\sin (2\delta) \, \vect{e} \Big) \,. 
\end{equation}
Again, the induced translational velocity of a freely moving drop subject to this net force follows as $\vect{V}_\mathrm{Drop}^\mathrm{D} = \mu \vect{F}_\mathrm{Drop}^\mathrm{D}$ and can thus be expressed as
\begin{equation}
	\vect{V}_\mathrm{Drop}^\mathrm{D} = - \frac{\alpha R \Lambda }{2+\Lambda} 
	\Big( \left( 3-\cos(2\delta) \right) \vect{d} - 3\sin (2\delta) \, \vect{e} \Big) \,. 
\end{equation}
Altogether, the total flow field resulting from a force-dipole acting inside a freely moving drop is obtained by superimposing the dipolar flow field inside a stationary drop derived above and the flow field induced by a drop translating with velocity~$-\vect{V}_\mathrm{Drop}^\mathrm{D}$ provided by the Hadamard-Rybczynski solution [\textit{c.f.}\ Eqs.~\eqref{hadamardOut} and~\eqref{hadamardIns}].
Again, for a rigid cavity and a surfactant-covered drop the total hydrodynamic force vanishes because $a_1^\parallel = a_1^\perp = 0$.

In analogy to the flow fields caused by a Stokeslet presented above, we now illustrate the flow induced by a force dipole. 
Figure~\ref{Dipo} shows corresponding results for a stiff spherical confinement of $\Lambda=0$  (left column) and for the case of $\Lambda=1/2$ (right column) for a clean freely moving drop in the absence of a surfactant. 
By varying the position of the force dipole inside the drop, we can control the additional recirculation zones appearing in the exterior fluid. The flow fields generally lose the axial symmetry, except for when the dipole is oriented radially.  Similarly, in Fig.~\ref{Stresslet}, we present related results caused by a pure stresslet.

Adding a surfactant significantly changes the observed dynamics. 
In Fig.~\ref{Dipo-Stresslet-Surfactant}, we present the flow fields caused by a dipole (left column) and a stresslet (right column) for various values of~$w$.
Increasing the shear viscosity of the surfactant leads to a `stiffening' that drastically reduces the effect of the singularity on the exterior flow. 
The exterior region consists of circular streamlines similar to those induced by rigid-body rotation.

\section{Swimmer dynamics}\label{sec:SwimmerDynamics}

We now analyze the effect of the drop on the dynamics of an active swimming microorganism encapsulated on the inside.
To this end, we decompose in the usual way the generated flow field into a bulk contribution given by Eq.~\eqref{vD-bulk} in addition to the correction due to the presence of the confining drop.
For a clean drop, an additional contribution has to be considered to account for the free motion of the drop.

Here, we model the swimming microorganism as a prolate spheroidal particle of aspect ratio~$\sigma$.
The latter is defined as the ratio of major to minor semi-axes of the spheroid.
For instance, the aspect ratio of the bacterium \textit{Bacillus subtilis}~\cite{ryan13} has been measured experimentally to be about~$\sigma = 4$.

The induced translational and rotational velocities resulting from the fluid-mediated hydrodynamic interactions between the microswimmer and the surface of the drop are provided by Fax\'{e}n's laws as~\cite{spagnolie12, spagnolie15, Mathijssen2016, daddi19pre}
\begin{subequations}
	\begin{align}
		\vect{v}^\mathrm{HI} &= \left. \vect{v}_\mathrm{D}^* (\vect{x}) \right|_{\vect{x} = \vect{x}_2} \, , \\[5pt]
		\boldsymbol{\Omega}^\mathrm{HI} &= 
		\tfrac{1}{2} \, \boldNabla \left. \times \vect{v}_\mathrm{D}^* (\vect{x}) \right|_{\vect{x} = \vect{x}_2}
		+ \Gamma \vect{q} \times \left. \vect{E}_\mathrm{D}^* (\vect{x}) \right|_{\vect{x} = \vect{x}_2} \cdot \vect{q} \, ,  
	\end{align}
\end{subequations}
where we have restricted these expressions to the leading order in the swimmer size.
Here, $\vect{v}_\mathrm{D}^*$ denotes the image dipole flow field inside a freely moving drop.
In addition, $\vect{E}_\mathrm{D}^*  = \left( \boldNabla \vect{v}_\mathrm{D}^* + \left(\boldNabla \vect{v}_\mathrm{D}^*\right)^\top \right)/2$ denotes the symmetric rate-of-strain tensor associated with the image force dipole, and $\top$ represents the transposition operation.
In addition, $\Gamma = \left( \sigma^2-1 \right)/\left( \sigma^2+1 \right) \in [0,1)$ is a shape factor, where $\Gamma = 0$ holds for a spherical particle and $\Gamma \to 1$ for a needle-like particle of a significantly pronounced aspect ratio.

Then, the induced translational velocity of the swimmer can be written as
\begin{equation}
	\vect{v}^\mathrm{HI} =  -\alpha 
	\Big( \big( V_1 + V_2 \cos \left(2\delta\right) \big) \vect{d} + V_3 \,   \sin \left(2\delta\right) \, \vect{e} \Big) \, , 
	\label{V_HI}
\end{equation}
where for a clean drop
{\allowdisplaybreaks
\begin{subequations}\label{V123clean}
	\begin{align}
		V_1	&= v_1 - \frac{\left(3-\Lambda\right)R}{4\left(1-R^2\right)^2} \, , \\
		V_2 &= v_2 + \frac{3\left(3-\Lambda\right)R}{4\left(1-R^2\right)^2} \, , \\
		V_3	&= v_3 + \sum_{n=1}^{\infty} \frac{3\left(1-\Lambda\right)\left(2n+1\right)\left(\Lambda-n-2\right)}{4 \left(n+2-3\Lambda \right)} \, R^{2n-1} \, , \label{V3clean}
	\end{align}
\end{subequations}
}where 
\begin{equation}
	-v_1 = 3 v_2 = 3LR \left(1-R^2\right) , \quad
	v_3 = 3L R \left(1-2R^2\right)
\end{equation}
are additional contributions required to account for the free motion of the drop, with
\begin{equation}
	L = \frac{\Lambda \left(1-\Lambda\right)}{2 \left(2+\Lambda\right)} \, .
\end{equation}
For a surfactant-covered drop, we obtain
{\allowdisplaybreaks
\begin{subequations}
	\begin{align}
		V_1 &= -\frac{V_2}{3} = -\frac{3R}{4\left(1-R^2\right)^2} \label{V12surfactant} \, , \\
		V_3 &= \sum_{n=1}^{\infty}\frac{u_n}{s_n} \, R^{2n-1} \, , \label{V3surfactant}
	\end{align}
\end{subequations}
}where we have defined
\begin{align}
	u_n &= -(2n+1) \big( 3wn^2 + \left(3+\lambda+3w\right)n +6-\lambda-6w \big) \, , \notag \\
	s_n &= 4 \big( wn^2 + \left(1+\lambda+w\right)n+2-\lambda-2w \big)  \, . \notag 
\end{align}

\begin{center}
	\begin{table*}
		\def\arraystretch{2.5}%
		\begin{tabular}{|c|c|c|c|c|}
			\hline
			$\Lambda$ & $V_3-v_3$ & $\Omega_1-\omega_1$ & $\Omega_2-\omega_2$ & $\Omega_3-\omega_3$ \\
			\hline\hline
			$0$ & $-\cfrac{3}{4} \cfrac{R \left( 3 - R^2 \right)}{\left( 1-R^2\right)^2}$ & $\cfrac{3}{4} \cfrac{R^2 \left( 3-R^2 \right)}{\left( 1-R^2 \right)^3}$
			& $-\cfrac{3}{32} \cfrac{R^4 \left( 5-R^2 \right)}{\left( 1-R^2 \right)^3}$
			& $\cfrac{3}{32} \cfrac{R^2 \left(16-5R^2+R^4\right)}{\left( 1-R^2 \right)^3}$ \\[5pt]
			\hline
			~~~~$\cfrac{1}{2}$~~~~ & ~~~~$-\cfrac{3}{8} \cfrac{R \left( 5 - 3R^2 \right)}{\left( 1-R^2\right)^2}$~~~~ & ~~~~$\cfrac{3}{4} \cfrac{R^2 \left( 3-R^2 \right)}{\left( 1-R^2 \right)^3}$~~~~
			& ~~~~$-\cfrac{3}{64} \cfrac{R^4 \left( 11 + R^2 \right)}{\left( 1-R^2 \right)^3}$~~~~ &
			~~~~$\cfrac{3}{64} \cfrac{R^2 \left( 24-3R^2-R^4 \right)}{\left( 1-R^2 \right)^3}$~~~~ \\[5pt]
			\hline
			$1$ & $0$ & $\cfrac{2}{\left( 1-R^2 \right)^3}$ & $-\cfrac{3}{8} \cfrac{R^4 \left( 3-R^2 \right)}{{\left( 1-R^2 \right)^3}}$ & $\cfrac{3}{8} \cfrac{R^2 \left( 4-3R^2+R^4 \right)}{{\left( 1-R^2 \right)^3}}$ \\[5pt]
			\hline\hline
		\end{tabular}
		\caption{Expressions of the infinite sums for a clean drop, given in Eqs.~\eqref{V123clean} and \eqref{Omega123clean} in terms of polynomial fractions for $\Lambda = 0$ (rigid-cavity limit), $\Lambda = 1/2$ (equal viscosities of the inner and outer fluids), and $\Lambda = 1$ (infinitely small outer viscosity).}
		\label{Tab:ExpressionSums}
		\vspace{0.5cm}
		\def\arraystretch{2.25}%
		\begin{tabular}{|c|c|c|c|}
			\hline
			$\vect{f} $ & $\left. \vect{f} \right|_{\vect{x} = \vect{x}_2 } $ & $ \left. \boldNabla \times \vect{f} \right|_{\vect{x} = \vect{x}_2 } $ & $ \vect{q} \times \left. \frac{1}{2} \left( \boldNabla \vect{f} + \left( \boldNabla \vect{f} \right)^\top  \right) \right|_{\vect{x} = \vect{x}_2 } \cdot \vect{q} $ \\
			\hline\hline
			$\vect{S}_n^{(1)} $ & $-\tfrac{1}{2} n (n+1) R^{n+1} \, \vect{d} $ & $\vect{0}$ & $\tfrac{3}{8} n \left(n+1\right)^2 R^n \sin (2\delta) \, \vect{t}$ \\
			\hline
			$\vect{S}_n^{(2)} $ & $-n R^{n-1} \, \vect{d}$ & $\vect{0}$ & $\tfrac{3}{4} n (n-1) R^{n-2} \sin (2\delta) \, \vect{t}$ \\
			\hline\hline
			$\vect{Q}_n^{(1)} $ & ~$-\frac{1}{4} \left(n+1\right)(n+3) R^{n+1} \, \vect{e} $~ & ~$-\tfrac{1}{2}  \left(n+1\right)  \left(2n+3\right) R^n \, \vect{t}$~ & $\tfrac{1}{4} \, n(n+1)(n+2) R^n \cos (2\delta ) \, \vect{t} $ \\
			\hline
			$\vect{Q}_n^{(2)} $ & $-\frac{1}{2} \left(n+1\right) R^{n-1} \, \vect{e}$ & $ \vect{0}$ & $\tfrac{1}{2} \left(n^2-1\right) R^{n-2} \cos (2\delta) \, \vect{t}$ \\
			\hline
			$\vect{Q}_n^{(3)} $ & $(n-1) R^{n-2} \, \vect{e}$ & $\tfrac{1}{2} \left(n-1\right) \left(n-2\right) R^{n-3} \, \vect{t} $ & $-\tfrac{3}{4} \left(n-1\right)(n-2) R^{n-3} \cos(2\delta) \, \vect{t}$ \\
			\hline\hline
			$\vect{W}_n^{(1)} $ & $-\tfrac{1}{4} n (n+1)^2 R^{n+1} \, \vect{d}$ & $\vect{0}$ & ~$\tfrac{1}{32} (n+1) \left( 7n^3+16n^2+7n-6 \right) R^n \sin(2\delta) \, \vect{t}$~ \\
			\hline
			$\vect{W}_n^{(2)} $ & $-\tfrac{1}{2} n (n+1) R^{n-1} \, \vect{d} $ & $\vect{0}$ & $\tfrac{1}{16} \left(n^2-1\right)(7n+2) R^{n-2} \sin(2\delta) \, \vect{t}$ \\
			\hline
			$\vect{W}_n^{(3)} $ & $\tfrac{1}{2} n (n-1) R^{n-2} \, \vect{d} $ & $\vect{0}$ & $-\tfrac{1}{2} n (n-1)(n-2) R^{n-3} \sin(2\delta) \, \vect{t}$ \\
			\hline\hline
		\end{tabular}
		\caption{Summary of the basic operations required for the calculations of the translational and rotational velocities given by Eqs.~\eqref{V_HI} and \eqref{Omega_HI}, respectively, for a surfactant-free, clean drop.}
		\label{Tab:Operations}
	\end{table*}
\end{center}

The induced rotational velocity due to hydrodynamic interactions with the surface of the drop can be cast in the form
\begin{equation}
	\boldsymbol{\Omega}^\mathrm{HI} = -\alpha \Big( \Omega_1 + \Gamma \big( \Omega_2 \cos(2\delta)+\Omega_3 \big) \Big) \sin(2\delta) \, \vect{t} \, ,  \label{Omega_HI}
\end{equation}
where, again, $\vect{t} = \vect{e} \times \vect{d}$.
For a clean drop, we find 
\begin{widetext}
\begin{subequations}\label{Omega123clean}
	\begin{align}
		\Omega_1 &= \omega_1 + \frac{3}{4} \sum_{n=1}^{\infty} 
		\frac{\left(n^2-1\right) \left( 2\Lambda^2-4\Lambda+2+n \right)}{ n+2-3\Lambda } \, R^{2n-2} \, , \\
		\Omega_2 &= \omega_2 +
		\frac{3 \left(3\Lambda - 1\right)}{32}
		- \frac{3}{32} \sum_{n=1}^{\infty} 
		\frac{(n-2) \left( 2(1+\Lambda)n^2+ \left(6\Lambda^2-13\Lambda+3\right)n - \left(1-\Lambda\right) \left(2+3\Lambda\right) \right)}{ n+2-3\Lambda } \, R^{2n-2} \, , \\
		\Omega_3 &= \omega_3 + \frac{3 \left(3\Lambda - 1\right)}{32}
		+ \frac{3}{32} \sum_{n=1}^{\infty} 
		\frac{ 2(3-\Lambda)n^3 + (2\Lambda-1)(5\Lambda-9)n^2 
		- (1-\Lambda) \left( (\Lambda+8)n + 2 (2-\Lambda) \right) }{ n+2-3\Lambda } \, R^{2n-2} \, ,
	\end{align}
\end{subequations}
where $\omega_1 = -15R^2L/2$, $\omega_2 = -9R^2L/2$, and $\omega_3 = 6R^2L$ are contributions accounting for the free motion of the drop.
For a surfactant-covered drop, we obtain 
\begin{subequations}\label{Omega123Surfactant}
	\begin{align}
		\Omega_1 &= \sum_{n=1}^{\infty}
		\frac{3wn^4+ \left(3+5\lambda+3w\right)n^3 + \left(6-2\lambda-9w\right)n^2 - \left(3+5\lambda+3w\right)n - 6+2\lambda+6w}{ 4 \left( wn^2 + \left(1+\lambda+w\right)n+2-\lambda-2w \right) } \, R^{2n-2} \, , \\
		\Omega_2 &= -\frac{3}{32} - \frac{3}{32}  \sum_{n=1}^{\infty} 
		\frac{2w n^4 + \left(2+6\lambda-3w\right)n^3 - \left(1+17\lambda+7w\right)n^2 + \left(9\lambda+12w-8\right)n + 4 + 2\lambda-4w}{ wn^2 + \left(1+\lambda+w\right)n+2-\lambda-2w} \, R^{2n-2} \, , \\
		\Omega_3 &= -\frac{3}{32} + \sum_{n=1}^{\infty}
		\frac{18w n^4 + \left(18+22\lambda+9w\right)n^3 + \left(27-21\lambda-51w\right)n^2 - \left( 24+3\lambda-12w \right)n - 12+2\lambda+12w}{ 32 \left( wn^2 + \left(1+\lambda+w\right)n+2-\lambda-2w \right)} \, R^{2n-2}\, .
	\end{align}
\end{subequations}
\end{widetext}
In particular, the induced translational and rotational swimming velocities inside a rigid spherical cavity are recovered when taking in Eqs.~\eqref{Omega123clean} and~\eqref{Omega123Surfactant} the limit $\lambda \to 0$.

\begin{figure}
	\includegraphics[scale=1]{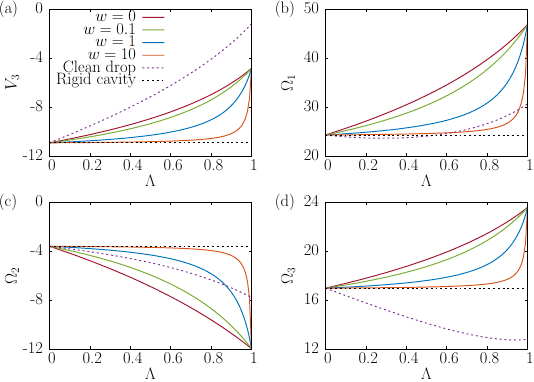}
	\caption{(Color online) Variation of the component~$V_3$ of the induced swimming velocity (a) and $\Omega_i$, $i \in \{1,2,3\}$, associated with the rotational swimming velocity [(b) -- (d)] inside a clean drop (magenta dashed line) and surfactant-covered drops (solid lines) for various values of scaled interfacial viscosity~$w$. The presence of a surfactant strongly alters the observed dynamics, particularly by enhancing reorientation.
	Here, we set $R=4/5$.}
	\label{Fig:SwimmingVelocities}
\end{figure}

\begin{figure}
	\includegraphics[scale=1]{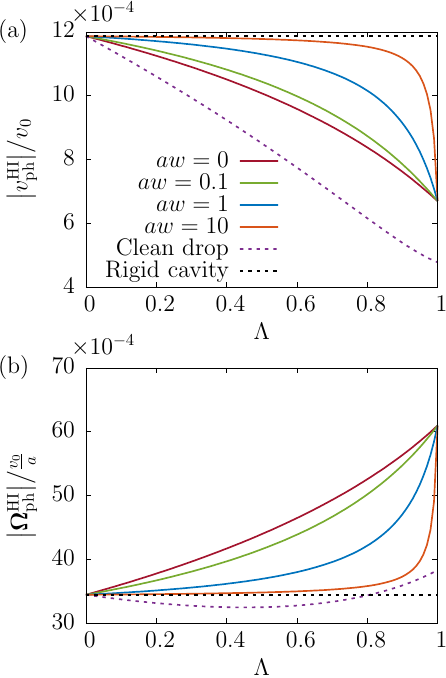}
	\caption{(Color online) Variations of the magnitude of the induced translational (a) and rotational (b) velocities for an \textit{E. coli} bacterium of aspect ratio $\sigma = 2$. 
	Here, we set $R=4/5$ and $\delta = \pi/4$.
	For the other parameters, see main text.}
	\label{Fig:SwimmingVelocities_ecoli}
\end{figure}

It is worth noting that the infinite series appearing in Eq.~\eqref{V3clean} providing the velocity~$V_3$ for a clean drop can be expressed in terms of Hurwitz-Lerch transcendent and Gauss hypergeometric functions~\cite{abramowitz72}.
However, the sum representation is more convenient for computational purposes.
For a clean drop, for $\Lambda = 0$ (corresponding to the rigid-cavity limit), for $\Lambda = 1/2$ (corresponding to equal viscosities of the inner and outer fluids), and for $\Lambda=1$ (corresponding to an infinitely small outer viscosity), the infinite sum can be expressed in terms of polynomial fractions as summarized in Tab.~\ref{Tab:ExpressionSums}.
For the sake of clarity, we summarize in Tab.~\ref{Tab:Operations} the basic operations that have been used to calculate the translational and rotational velocities stated by Eqs.~\eqref{V_HI} and \eqref{Omega_HI}, respectively.
In Appendix.~\ref{appendix:2}, we discuss the convergence properties of these series functions and estimate the number of terms required for their evaluation up to a given precision.

The addition of a surfactant increases the complexity of the solution.
The magnitudes of the velocities~$V_1$ and~$V_2$ for a surfactant-covered drop given by Eq.~\eqref{V12surfactant} are independent of~$\Lambda$ and are generally larger than those for a clean drop given by Eqs.~\eqref{V123clean}.
In Fig.~\ref{Fig:SwimmingVelocities}, we present the $\Lambda$-dependence of the components $V_3$ and $\Omega_i$, $i \in \{1,2,3\}$, for different values of the surface viscosity ratio $w$. 
The induced translational swimming velocity~$V_3$ and the rotation rates increase monotonically from the rigid-cavity limit $(\Lambda=0)$ to the infinitely large viscosity contrast $(\Lambda=1)$.  
The presence of a surfactant strongly alters the dynamics of the encapsulated swimmer by enhancing its reorientation when compared to the situation of a clean drop.

Finally, we exploit our results to estimate the velocity and rotation rates for real biological microswimmers confined by the spherical drop.
As an example, we chose the bacterium \textit{E. coli}, which provides a frequently studied example system to unravel the physics of microswimmers~\cite{berg73, berg2008coli, schwarz2016escherichia}.
We recall that throughout the article, we have rescaled all length by the radius $a$ of the drop.
In the following, we use the subscript  ``ph'' to denote non-scaled quantities in physical units, which implies $\alpha_\mathrm{ph} = a^3 \alpha$, $\vect{v}^\mathrm{HI}_\mathrm{ph} = a \, \vect{v}^\mathrm{HI}$, and 
$\boldsymbol{\Omega}^\mathrm{HI}_\mathrm{ph} = \boldsymbol{\Omega}^\mathrm{HI}$.
The functions $V_i$ and~$\Omega_i$, $i \in \{1,2,3\}$, are dimensionless quantities.

In a bulk Newtonian fluid of dynamic viscosity $\eta = 10^{-3}~$~Pa$\cdot$s, \textit{E. coli} bacteria swim with an average velocity of $v_0 \approx 20~\mu$m/s.
By approximating the flagella thrust forces as $f \approx 0.5~$pN~\cite{chattopadhyay2006swimming, darnton2007torque} and the inter-dipole distance as $\ell \approx 1~\mu$m~\cite{berke08}, the resulting force dipolar coefficient is estimated as $\alpha_\mathrm{ph} = f\ell/\left( 8\pi\eta \right) \approx 20~\mu$m$^3$/s.
Rescaling the induced translational velocity by~$v_0$ and the rotation rate by $v_0/a$, it follows from Eqs.~\eqref{V_HI} and \eqref{Omega_HI} that
{\allowdisplaybreaks
	\begin{subequations}\label{Velocities_ecoli}
	\begin{align}
		\frac{\left| \vect{v}_\mathrm{ph}^\mathrm{HI} \right|}{v_0 }
		&= \Xi \, \Big( \big( V_1 + V_2 \cos \left(2\delta\right) \big)^2 + V_3^2 \,   \sin^{2} (2\delta) \Big)^{\frac{1}{2}} \, , \label{V_HI_ecoli} \\
		\frac{\left| \boldsymbol{\Omega}_\mathrm{ph}^\mathrm{HI} \right|}{v_0/a} &= \Xi \left| \Big( \Omega_1 + \Gamma \big( \Omega_2 \cos(2\delta)+\Omega_3 \big) \Big) \sin(2\delta) \right| \, , \label{Omega_HI_ecoli} 
	\end{align}
\end{subequations}
}where $\Xi := \alpha_\mathrm{ph} / \left( a^2 v_0 \right) = 10^{-4}$.

In Fig.~\ref{Fig:SwimmingVelocities_ecoli}, we present the variation of the magnitude of the rescaled swimming velocities as stated by Eqs.~\eqref{Velocities_ecoli} versus the parameter~$\Lambda$ for various values of the surface to external bulk viscosity ratio~$w$.
Here, we consider a spherical viscous drop of radius $a = 0.1~$mm.
To limit the parameter space, we assume that the bacterium has an orientation angle $\delta = \pi/4$ and an aspect ratio $\sigma = 2$ $(\Gamma = 3/5)$ \cite{berke08}.
Compared to a microswimmer in a clean drop, we observe that the presence of a surfactant enhances the magnitude of the induced translational velocity as well as the rotation rate.
For increasing $w$, the magnitude of the induced translational velocity approaches the maximal value found for a swimmer inside a rigid cavity. 
Whether the induced translational velocity impedes or supports the translation of the microswimmer depends, in the end, on the orientation angle $\delta$.
In contrast to that, the effect of the surfactant on the rotation rate is weakened with increasing $w$ and is the most severe in the case of vanishing shear viscosity (i.e., $w=0$), for which particularly the incompressibility of the surfactant on the surface of the drop distinguishes the situation from that of a clean drop. 
Since $| \vect{v}^\mathrm{HI}_\mathrm{ph} | \sim a^{-2}$ and $| \boldsymbol{\Omega}^\mathrm{HI}_\mathrm{ph} | \sim a^{-3}$, the confinement effect on the induced swimming velocities and rotation rates becomes more important upon decreasing the size of the drop.

\section{Conclusions}\label{sec:Conclusions}

Stokes flows in complex and confined geometries have significant relevance for a variety of applications in industrial and biological systems. In this context, drops of particular importance, because a number of microfluidic realizations exploits their potential for trapping active or passive particles and biological material, including proteins, biopolymers, and microswimmers. Understanding the dynamics inside these micro-containers requires an adequate description of the flow generated by the enclosed matter.

In this contribution, we have developed analytical expressions for the lowest-order flow singularities, namely the Stokeslet and force dipole, enclosed inside a liquid drop surrounded by a fluid environment. 
We have explored the flow structure for arbitrary viscosity contrast between the spherical drop and the suspending fluid. 
First, we have provided our results both for the case when the drop is clean, and thus tangential stresses are continuous across the boundary. 
Second, we have analyzed the effect of the presence of a homogeneously distributed, incompressible surfactant on the surface of the drop on the resulting flow fields. 
To model the surfactant, we have employed the Boussinesq-Scriven constitutive law, with the surfactant characterized by an interfacial shear viscosity. 
Using spherical harmonic expansion techniques, we have been able to determine the flow fields in each case and present them for a varying interior/exterior fluid viscosity ratio and also for different values of the surfactant shear viscosity.

Having derived the image flows in each case, we have further discussed the effective forces exerted on the surface of the drop due to the presence of the enclosed point singularities.
On our way, this was necessary to render the drop moving freely. 
Next, we have focused our discussion on the case of drops with entrapped microswimmers and found the resulting translational and rotational velocities of force- and torque-free swimmers inside such spherical confinements. To this end, we have used the Fax\'{e}n relations and modeled the swimmer as a prolate spheroid. 
We have found that the presence of the surfactant tends to enhance the rotation rate of the encapsulated swimmer.

The results derived in this paper constitute a step towards understanding the complex dynamics resulting from hydrodynamic interactions in a confined and complex environment. 
The minimal model proposed here for the interpretation of any experimentally observed motion of active or passive particles can be directly employed to describe the dynamics observed in flows both internal and external to the drop, \textit{e.g.}\ in colloidal suspension of microdrops and microfluidic diagnostic devices.

\begin{acknowledgments}
	We would like to thank B.\ Ubbo Felderhof and Shubhadeep Mandal for various stimulating discussions. 
	H.L. and A.M.M. gratefully acknowledge support from the Deutsche Forschungsgemeinschaft (DFG) through the priority program SPP 1726 on microswimmers, grant nos.\ LO~418/16 and ME~3571/2. 
	A.D.M.I. thanks the DFG for support through the project DA~2107/1-1.
	A.M.A acknowledges support from National Science Foundation (CBET-1700961). V.A.S acknowledges support from Bisland Dissertation Fellowship.
	A.J.T.M.M. acknowledges funding from the Human Frontier Science Program (fellowship LT001670/2017), and the United States Department of Agriculture (USDA-NIFA AFRI grant 2019-06706).	
	F.G.L.\ acknowledges Millennium Nucleus Physics of Active Matter of ANID (Agencia Nacional de Investigaci\'on y Desarrollo).
\end{acknowledgments}

\section*{Author contribution statement}
A.D.M.I. conceived the study and prepared the figures. 
A.R.S. and A.D.M.I. carried out the analytical calculations.
V.A.S., M.L., and A.D.M.I. drafted the manuscript. 
All authors discussed and interpreted the results, edited the text, and finalized the manuscript.

\appendix

\begin{widetext}
\section{Determination of the series coefficients}\label{appendix:1}

In this Appendix, we provide the resulting equations for the boundary conditions stated by Eqs.~\eqref{boundaryCond1} and \eqref{boundaryCond2clean} for clean drop and by Eqs.~\eqref{boundaryCond1} and \eqref{boundaryCond2surfactant} for a surfactant-covered drop.

\subsection{Axisymmetric Stokeslet}

As already mentioned, the solution of the axisymmetric flow field induced by a point force inside a surfactant-covered drop is identical to that inside a rigid cavity for which $\lambda \to 0$ (or alternatively $\Lambda\to 0$).
Thus, in the axisymmetric case, we will provide in the following the resulting equations for the boundary conditions for a clean viscous drop only.

By applying the boundary conditions of vanishing radial velocity at the surface of the drop, given by Eq.~\eqref{VanishingRadial}, and using the fact that $\boldNabla \varphi_n$ and $\R \varphi_n$ form a set of independent vector harmonics, we find
\begin{equation}
	\frac{n(n+1)}{2} \, A_n^\parallel  + nB_n^\parallel = \frac{n(n+1)}{2n-1} \, R^{n-1} - \frac{n(n+1)}{2n+3} \, R^{n+1} \, , \qquad 
	\frac{n(n+1)}{2} \, a_n^\parallel - (n+1) b_n^\parallel = 0 \, . \label{SysAxi1}
\end{equation}

In addition, the continuity of the tangential components of the velocity and stress vector fields, respectively given by Eqs.~\eqref{ContVelo} and~\eqref{boundaryCond2clean}, leads to two additional equations
\begin{subequations}\label{SysAxi2}
	\begin{align}
		-\frac{n+3}{2} \, A_n^\parallel  - B_n^\parallel -\frac{n-2}{2} \, a_n^\parallel + b_n^\parallel  &= \frac{n-2}{2n-1} \, R^{n-1} - \frac{n}{2n+3} \, R^{n+1} \, , \\
		 \frac{n(n+3)}{2} \, A_n^\parallel + (n-2) B_n^\parallel -\frac{(n+1)(n-2)}{2\lambda} \, a_n^\parallel + \frac{n+3}{\lambda} \, b_n^\parallel &=  \frac{(n+1)(n-2)}{2n-1} \, R^{n-1} - \frac{n(n+3)}{2n+3} \, R^{n+1} \, .
	\end{align}
\end{subequations}
Equations~\eqref{SysAxi1} and \eqref{SysAxi2} form a linear system of equations, amenable to resolution using the standard substitution technique. 
From here, we obtain the expressions of the series coefficients $\{A_n^\parallel , B_n^\parallel \}$ and $\{a_n^\parallel, b_n^\parallel\}$ associated with the solution for the flow field inside and outside the drop, respectively, see Eqs.~\eqref{Coeffs_Parallel} of the main text.

\subsection{Transverse Stokeslet}

Applying the boundary condition of vanishing radial velocity field at the surface of the drop, as given by Eq.~\eqref{VanishingRadial}, yields
\begin{equation}
	\frac{n+1}{2} \, A_n^\perp + B_n^\perp - C_{n+1}^\perp = \frac{n+1}{2n-1} \, R^{n-1} - \frac{n+3}{2n+3} \, R^{n+1} \, ,  \qquad 
	-\frac{n}{2} \, a_n^\perp + b_n^\perp - c_{n+1}^\perp = 0 \, . \label{SysAsymm1}
\end{equation}

In addition, the continuity of the tangential components of the velocity vector field, as given by Eq.~\eqref{ContVelo}, implies
\begin{subequations}\label{continuityTanVelo}
	\begin{align}
		\frac{n+1}{n+2} \, C_{p+3}^\perp + c_{n+1}^\perp &= -\frac{2R^{n+1}}{n+2} \, , \\
		\frac{n+3}{2n} \, A_n^\perp + \frac{B_n^\perp}{n} - \frac{C_{n+1}^\perp}{n} + \frac{C_{n+3}^\perp}{n+2} - \frac{(n-2) \, a_n^\perp}{2(n+1)}  + \frac{b_n^\perp}{n+1} &= \left( -\frac{n-2}{n(2n-1)} + \frac{n R^{2}}{(n+2)(2n+3)} \right) R^{n-1} \, .
	\end{align}
\end{subequations}

On the one hand, for a clean drop, the continuity of the tangential hydrodynamic stresses stated by Eq.~\eqref{boundaryCond2clean} yields
\begin{subequations}\label{continuityTanStress}
	\begin{align}
		\frac{n(n+1)}{n+2} \, C_{n+3}^\perp - \frac{n+3}{\lambda} \, c_{n+1}^\perp &= \frac{2(n+3)}{n+2} \, R^{n+1} \, , \\
		\frac{n+3}{2} \, A_n^\perp + \frac{n-2}{n} \left( B_n^\perp - C_{n+1}^\perp \right) + \frac{n C_{n+3}^\perp}{n+2} +   \frac{n-2}{2\lambda} \, a_n^\perp - \frac{n+3}{\lambda(n+1)} \, b_n^\perp &= \left(  \frac{(n+1)(n-2)}{n(2n-1)}  - \frac{n(n+3) R^{2}}{(2n+3)(n+2)} \right) R^{n-1} \, .
	\end{align}
\end{subequations}

Next we solve Eqs.~\eqref{SysAsymm1} through \eqref{continuityTanStress} for the series coefficients $\{ A_n^\perp, B_n^\perp, C_n^\perp \}$ and $\{ a_n^\perp, b_n^\perp, c_n^\perp \}$ associated with the flow fields inside and outside the drop, respectively.
Explicit expressions of these coefficients are given in Eq.~\eqref{Coeffs_Perp_Clean}.

On the other hand, for a surfactant-covered drop, Eqs.~\eqref{boundaryCond2surfactant} representing the incompressibility of the in-plane surfactant flow and the discontinuity of the tangential hydrodynamic stresses, lead to
\begin{subequations}\label{IncompressUndBoussinesq}
	\begin{align}
		\left( \frac{n}{2}-1 \right) a_n^\perp - b_n^\perp + c_{n+1}^\perp &= 0 \, , \\
		n(n+1) \left( wn+\lambda+3w \right) C_{n+3}
		-(n+2)(n+3) c_{n+1} &= 2(n+3) \left(\lambda-wn\right) R^{n+1} \, . \label{BoussinesqProjected}
	\end{align}
\end{subequations}
It is worth noting that Eq.~\eqref{BoussinesqProjected} is obtained upon operating $\vect{e}_r \cdot \boldsymbol{\nabla}_\mathrm{S} \times $ on both sides of Eq.~\eqref{Boussinesq} to eliminate the term $\boldsymbol{\nabla}_\mathrm{S} \gamma$. 

The expressions of the series coefficients follow forthwith upon solving the linear system of equations composed of Eqs.~\eqref{VanishingRadial}, \eqref{continuityTanVelo}, and \eqref{IncompressUndBoussinesq} to yield the expressions given by Eqs.~\eqref{Coeffs_Perp_Surfactant} of the main body of the paper. 
\end{widetext}

\section{Convergence and estimation of the number of terms required for the evaluation of infinite series functions}\label{appendix:2}

In this Appendix, we discuss the convergence of the series functions for the induced translational and rotational swimming velocities given by Eqs.~\eqref{V3clean} and \eqref{Omega123clean} for a clean drop, and by Eqs.~\eqref{V3surfactant} and \eqref{Omega123Surfactant} for a surfactant-covered drop.

Let us denote by ${v_3}_n$ the general term of the infinite series for $V_3$ given for a clean drop Eq.~\eqref{V3clean}, \textit{i.e.}, $V_3 = \sum_{n=1}^{\infty} {v_3}_n$.
To test the convergence of the series, we define in the usual way the ratio $L = \lim\limits_{n\to\infty} \left| {v_3}_{n+1}/{v_3}_n \right| = R^2 < 1$ in rescaled units of length.
Therefore, the series is absolutely convergent~\cite{folland99}. 
Then, for $n \sim \infty$, we have the leading-order asymptotic behavior
\begin{equation}
	{v_3}_n = -\frac{3}{4} \left(1-\Lambda\right) \left(2n+1+4\Lambda\right)R^{2n-1}
	+ \mathcal{O} \left( \tfrac{R^{2n}}{n} \right) .
\end{equation}

To compute the infinite series at a given desired precision, we define the truncation error as
\begin{equation}
	\mathcal{E} \{V_3\} := \left| \sum_{n=N+1}^{\infty}  {v_3}_n \right| 
		 \simeq  \frac{3\left(1-\Lambda\right)N}{2\left(1-R^2\right)} \, R^{1+2N} \, .
\end{equation}
The number of terms required to achieve a certain precision~$\varepsilon$ can readily be obtained by solving numerically the inequality $\mathcal{E} (V_3) < \varepsilon$.

For a surfactant-covered drop, it can be shown that 
\begin{equation}
	\mathcal{E}  \{V_3\} \simeq  \frac{3N}{2\left(1-R^2\right)} \, R^{1+2N} \, .
\end{equation}

We proceed analogously for the series functions for the rotational velocity given for a clean drop by Eq.~\eqref{Omega123clean}.
Here, we obtain 
\begin{subequations}
	\begin{align}
		\mathcal{E} \{\Omega_1\} &\simeq \frac{3N^2}{4 \left(1-R^2\right)} \, R^{2N} \, , \\
		\mathcal{E} \{\Omega_2\} &\simeq \frac{3\left(1+\Lambda\right)N^2}{16 \left(1-R^2\right)} \, R^{2N} \, , \\
		\mathcal{E} \{\Omega_3\} &\simeq \frac{3\left(3-\Lambda\right)N^2}{16 \left(1-R^2\right)} \, R^{2N} \, . 
	\end{align}
\end{subequations}

Similarly, for a surfactant-covered drop, we obtain
\begin{equation}
	\mathcal{E} \{\Omega_1\} 
	\simeq 4\mathcal{E} \{\Omega_2\} 
	\simeq \frac{4}{3} \,  \mathcal{E} \{\Omega_3\} 
	\simeq \frac{3N^2}{4 \left(1-R^2\right)} \, R^{2N} \, .
\end{equation}

For instance, for $R=4/5$ about 30 -- 40 terms are required for $\varepsilon = 10^{-3}$ whereas about 40 -- 50 terms are required for $\varepsilon = 10^{-6}$.
The number of required terms increases quickly as $R \to 1$.

\input{main.bbl}
\end{document}

%% file: commands.tex
\newcommand{\vect}[1]{\boldsymbol{#1}}

\newcommand{\boldNabla}{\boldsymbol{\nabla}}

\newcommand{\G}{\mathcal{G}}
\newcommand{\R}{\vect{r}}
\newcommand{\Intd}{\mathrm{d }}

\newcommand{\X}{\vect{x}}

\newcommand{\vStok}{\vect{v}^{\mathrm{S}}}

\newcommand{\xOne}{\vect{x}_1}
\newcommand{\xTwo}{\vect{x}_2}

\newcommand{\vecD}{\vect{d}}

\newcommand{\infSumOne}{\sum_{n=1}^{\infty}}



%% file: main.bbl
\providecommand{\noopsort}[1]{}\providecommand{\singleletter}[1]{#1}%